\documentclass[preprintnumbers, floatfix,letterpaper,aps,prd,epsfig,nofootinbib,
twocolumn
]{revtex4-1}
\usepackage{bm,graphicx,dcolumn,epstopdf,epsf, latexsym,mathbbol, amssymb,amsmath,color,slashed, mathrsfs,mathcomp,simplewick, multirow}
\usepackage[center]{subfigure}
\usepackage{multirow}
\usepackage{makecell}
\usepackage[colorlinks,linkcolor=blue,citecolor=blue,urlcolor=blue]{hyperref}

\begin{document}
\allowdisplaybreaks
 \newcommand{\bq}{\begin{equation}}
 \newcommand{\eq}{\end{equation}}
 \newcommand{\bqn}{\begin{eqnarray}}
 \newcommand{\eqn}{\end{eqnarray}}
 \newcommand{\nb}{\nonumber}
 \newcommand{\lb}{\label}
 \newcommand{\f}{\frac}
 \newcommand{\p}{\partial}
\newcommand{\PRL}{Phys. Rev. Lett.}
\newcommand{\PLB}{Phys. Lett. B}
\newcommand{\PRD}{Phys. Rev. D}
\newcommand{\CQG}{Class. Quantum Grav.}
\newcommand{\JCAP}{J. Cosmol. Astropart. Phys.}
\newcommand{\JHEP}{J. High. Energy. Phys.}
\newcommand{\red}{\textcolor{red}}
%

\title{Gravitational-wave constraints on spatial covariant gravities}

\author{Tao Zhu${}^{a, b}$}
\email{zhut05@zjut.edu.cn; Corresponding author}

\author{Wen Zhao${}^{c, d}$}
\email{wzhao7@ustc.edu.cn}

\author{Anzhong Wang${}^{e}$}
 \email{anzhong$\_$wang@baylor.edu}

\affiliation{
${}^{a}$ Institute for theoretical physics and cosmology, Zhejiang University of Technology, Hangzhou, 310032, China \\
${}^{b}$ United Center for Gravitational Wave Physics (UCGWP), Zhejiang University of Technology, Hangzhou, 310032, China \\
${}^{c}$ CAS Key Laboratory for Research in Galaxies and Cosmology, Department of Astronomy, University of Science and Technology of China, Hefei 230026, China \\
${}^{d}$ School of Astronomy and Space Sciences, University of Science and Technology of China, Hefei, 230026, China \\
${}^{e}$ GCAP-CASPER, Physics Department, Baylor University, Waco, Texas 76798-7316, USA}

\date{\today}

\begin{abstract}

The direct discovery of gravitational waves (GWs) from the coalescence of compact binary components by the LIGO/Virgo/KAGRA Collaboration provides an unprecedented opportunity for exploring the underlying theory of gravity that drives the coalescence process in the strong and highly dynamical field regime of gravity.  In this paper, we consider the observational effects of spatial covariant gravities on the propagation of GWs in the cosmological background and obtain the observational constraints on coupling coefficients in the action of spatial covariant gravities from GW observations. We first decompose the GWs into the left- and right-hand circular polarization modes and derive the effects of the spatial covariant gravities on the propagation equation of GWs. We find that these effects can be divided into three classes: 1) frequency-independent effects on GW speed and friction, 2) parity-violating amplitude and velocity birefringences, and 3) a Lorentz-violating damping rate and dispersion of GWs. With these effects, we calculate the corresponding modified waveform of GWs generated by the coalescence of compact binaries. By comparing these new effects with the publicly available posterior samples or results from various tests of gravities with LIGO/Virgo/KAGRA data in the literature, we derive the observational constraints on coupling coefficients of the spatial covariant gravities. These results represent the most comprehensive constraints on the spatial covariant gravities in the literature.

\end{abstract}


\maketitle

\section{Introduction}
\renewcommand{\theequation}{1.\arabic{equation}} \setcounter{equation}{0}

The direct detection of gravitational waves (GWs) by LIGO/Virgo Collaboration has ushered in an entirely new era of gravitational-wave astronomy \cite{gw150914, gw170817, gw-other, LIGOScientific:2017ycc, gwtc1, gwtc2, LIGOScientific:2021djp}.  To date, the LIGO/Virgo/KAGRA Collaboration has announced the detection of more than 90 confident GW events in the Gravitational-Wave Transient Catalog (GWTC) \cite{gwtc1, gwtc2, LIGOScientific:2021djp}. These signals are produced by the coalescence of compact binaries, including binary black holes (BBH), binary neutron stars, and black hole-neutron star binaries. The GWs of these events, carrying valuable information about local spacetime properties of the compact binaries, allow us to explore the extreme gravity regime of spacetime, where the fields are strong, nonlinear, and highly dynamical. This has enabled a lot of model-independent tests of general relativity (GR) by the LIGO/Virgo/KAGRA Collaboration \cite{gw150914-testGR, gw170817-testGR, gw170817-speed, testGR_GWTC1, testGR_GWTC2, LIGOScientific:2021sio}. All of the tests to date have confirmed that GW data is consistent with the predictions of GR. 

With these substantial successes of GR,  the increasing number of detected GW events from the LIGO/Virgo/KAGRA Collaboration also provides a valuable window to explore, distinguish, or constrain any modified theories that exhibit deviations from GR. This has stimulated a lot of work on constraining different modified theories of gravity with GW data. In this paper, we consider a specific type of modified theory of gravity, the spatial covariant gravities \cite{Gao:2014soa, Gao:2014fra, Gao:2019lpz, Gao:2020yzr}, and test them with the current population of GW events. 

The spatial covariant gravities represent a series of alternative modified theories of GR, which break time diffeomorphism invariance but preserve the spatial one \cite{Gao:2014soa, Gao:2014fra}. This is very similar to the case of the Ho\v{r}ava-Lifshitz (HL) theory of quantum gravity \cite{Horava:2009uw, Horava:2010zj, Zhu:2011xe, Zhu:2011yu, wang2013,zhu2013, Wang:2017brl, Gao:2018znj}, in which the symmetry of the theory is broken from the general covariance down to the foliation-preserving diffeomorphisms. Such spatial covariance allows one to construct the action of the theory only in terms of spatial diffeomorphism invariants and study the effects of different terms. These new terms, which are absent in the Einstein-Hilbert action of GR, provide an efficient way to parametrize unknown high-energy physics effects on the low-energy scale. On the other hand,  the spatial covariant gravities can also represent a very general framework for describing a lot of scalar-tensor theories in unitary gauge \cite{Gao:2014soa, Gao:2020yzr, Joshi:2021azw, PGWs1}.  To our knowledge, a lot of scalar-tensor theories can be mapped to the spatial covariant framework by imposing the unitary gauge, including Horndeski theory, Chern-Simons modified gravity, Weyl gravity, ghost-free parity-violating gravities, $D\to 4$ Gauss-Bonnet gravity, Ho\v{r}ava-Lifshitz gravities, etc. (see details in Ref. \cite{Gao:2019liu}). 

One natural question now is whether the new terms beyond GR introduced in the spatial covariant gravities can lead to any observational effects in the current and/or forthcoming experiments and observations, so the spatial covariant gravities can be tested or constrained directly by observations. Such considerations have attracted a great deal of attention lately and several phenomenological implications of the spatial covariant gravities have already been investigated \cite{Gao:2019liu, Zhu:2022dfq, Gong:2021jgg, Zhu:2022caa, Hiramatsu:2022ahs, Iyonaga:2021yfv, Ganz:2022iiv}. The phenomenological implications for other theories that can also be described in the spatial covariant framework under certain conditions have also been extensively studied, see for example, Refs. \cite{Nilsson:2022mzq, llifshitz2, PGWs1, Wang:2012fi, Qiao:2019hkz, Zhu:2013fja, Li:2022grj} and reference therein. In particular, the effects of the spatial covariant gravities on the propagation of GWs in the cosmological background were previously explored in \cite{Gao:2019liu}. In addition, the imprints of the spatial covariant gravities on the primordial GWs was also calculated in detail recently in Ref. \cite{Zhu:2022dfq}. It was shown that the parity-violating terms in the gravitational action of the spatial covariant gravities can induce a nonzero circular polarization in the primordial GWs. The possible signatures of these parity-violating effects on the cosmic microwave background and the statistics of galaxy surveys were also briefly explored in Ref.~\cite{Zhu:2022dfq}.  

In this paper, we focus on the imprints of spatial covariant gravities on the propagation of GWs, produced by the coalescence of compact binaries, and their observational constraints with observational data of GW events from LIGO/Virgo/KAGRA Collaboration. Decomposing the GWs into the left- and right-hand circular polarization modes, we find that the equations of motion of GWs in spatial covariant gravities can be exactly mapped to the parametrized propagation equation proposed in Refs.~\cite{Zhao:2019xmm, Qiao:2019wsh}. Depending on different terms in the gravitational action of spatial covariant gravities, the new effects beyond GR can be fully characterized by four parameters: $\bar \nu$, $\bar \mu$, $\nu_A$, and $\mu_A$. The parameters $\nu_A$ and $\mu_A$ label the effects of the parity-violating terms in the spatial covariant gravities, and $\bar \nu$ and $\bar \mu$ describe the effects of other possible modifications that are not relevant to parity violation. The correspondences between different terms in the spatial covariant gravities and the four parameters are summarized in Table.~\ref{paramaters}. In addition, we present the expressions for these four parameters for a number of specific theories in Table \ref{specific_theories}.

Different parameters correspond to different effects on the propagation of GWs. These effects can be divided into three classes: 1) the frequency-independent effects which include the modification to the speed of GWs and GW friction, 2) the parity-violating effects which include the amplitude and velocity birefringences of GWs, and 3) the Lorentz-violating effects which include the modified damping rate and dispersion relation of GWs. We compare these new effects with existing observational samples or results of various tests of gravities with LIGO/Virgo/KAGRA data in the literature and derive the observational constraints on different terms in the spatial covariant gravities. Using these constraints, we also derive the corresponding bounds on the coupling coefficients of a number of specific theories in the Appendix. Our results are summarized in Table \ref{table4}. With the future ground- and space-based detector network, more and more GW events in a wider frequency range will be detected in the future, and we expect that the constraints on the spatial covariant gravities will be improved dramatically and a deeper understanding of the nature of gravity will be achieved.

This paper is organized as follows. In the next section, we present a brief review of the spatial covariant gravities, and in Sec. III we discuss the associated propagation of GWs in a homogeneous and isotropic cosmological background. In that section, we also map the different new effects on the propagation of GWs arising from the spatial covariant gravities into four parameters. In Sec. IV, we calculate the effects of the spatial covariant gravities on the speed, frictions, and waveform of GWs produced by the coalescence of compact binaries, and derive their observational constraints with observational data of GW events from the LIGO/Virgo/KAGRA Collaboration. A brief summary of our main results and some discussions are presented in Sec. V. We also present the detailed correspondences between the coupling constants in a number of specific theories and the coefficients in the spatial covariant gravities and derive the corresponding observational bounds on these specific theories in the Appendix.

Throughout this paper, the metric convention is chosen as $(-,+,+,+)$, and greek indices $(\mu,\nu,\cdot\cdot\cdot)$ run over
$0,1,2,3$ and latin indices $(i, \; j,\;k)$ run over $1, 2, 3$. We set the units to $c=\hbar=1$.

\section{spatial covariant gravities}
\renewcommand{\theequation}{2.\arabic{equation}} \setcounter{equation}{0}

In this section, we present a brief introduction of the construction of the spatial covariant gravities. Most of the expressions and results used here can be found in Refs. \cite{Gao:2014soa, Gao:2014fra, Zhu:2022dfq, Gao:2019liu} and references therein. 

The spatial covariant gravity is only invariant under the three-dimensional spatial diffeomorphism, which breaks the time diffeomorphism.  Therefore, the gravitational action of this type of theory can be constructed only in terms of spatial diffeomorphism invariants. In order to write down the gravitational action, it is convenient to write the metric of the spacetime in the Arnowitt-Deser-Misner (ADM) form \cite{ADM},
\bqn
ds^2 = - N^2 dt^2 + g_{ij} (dx^i + N^i dt)(dx^j + N^j dt),
\eqn
where $N$ is the lapse function, $g_{ij}$ is the three-dimensional spatial metric, and $N_i$ is the shift vector. With these ADM variables, the general action of the spatial covariant gravities can be written in the form,
\bqn\lb{action}
S = \int dt d^3 x N \sqrt{g} {\cal L}(N, g_{ij}, K_{ij}, R_{ij}, \nabla_i, \varepsilon_{ijk}),
\eqn
where $K_{ij}$ is the extrinsic curvature of $t=$const hypersurfaces,
\bqn
K_{ij} = \frac{1}{2N} \left(\partial_t g_{ij} - \nabla_i N_j -\nabla_j N_i\right), 
\eqn
$R_{ij}$ is the intrinsic curvature tensor, $\nabla^i$ is the spatial covariant derivative with respect to $g_{ij}$, and $\varepsilon_{ijk}=\sqrt{g} \epsilon_{ijk}$ is the spatial Levi-Civita tensor with $\epsilon_{ijk}$ being the total antisymmetric tensor. 

Normally, with the breaking of the time diffeomorphism, extra degrees of freedom are often added on top of the two tensorial degrees of freedom in GR. In particular, the spatial covariant gravity which has three dynamical degrees of freedom has been explored extensively \cite{Gao:2014fra}.  It was also shown that under two necessary and sufficient conditions, the spatial covariant gravities can have just two tensorial degrees of freedom and no propagating scalar mode \cite{Gao:2019twq}. One condition is the degenerate condition, which requires the lapse-extrinsic curvature sector of the Dirac matrix must be degenerate. Another condition is the consistent condition, which requires that the dimension of the phase space at each spacetime point must be even.  In Refs. \cite{Gao:2018znj, Gao:2019lpz}, the above action has also been extended by introducing $\dot N$ in the Lagrangian through $\frac{1}{N}(\dot N - N^i \nabla_i N)$. This term can only contribute to the scalar-type modes of GWs which are expected to be small, compared to the observed tensorial modes. For this reason, we will not consider it in this paper.

\begin{table*}
\caption{Building blocks of spatial covariant gravities up to fourth order in derivatives of $h_{ij}$, where $d_t$ and $d_s$ are the numbers of time and spatial derivatives respectively, and $d=d_t+d_s$ denotes the total number of time and spatial derivatives. Here $\omega_3(\Gamma)$ denotes the spatial gravitational Chern-Simons term, and $\omega_3(\Gamma) = \varepsilon^{ijk} (\Gamma^m_{jl} \partial_j \Gamma^l_{km} + \frac{2}{3} \Gamma^n_{il} \Gamma^l_{jm} \Gamma^m_{kn})$ with $\Gamma^k_{ij} = \frac{1}{2} g^{km} (\partial_j g_{mj} + \partial_j g_{ij} - \partial_m g_{ij})$ are the spatial Christoffel symbols. The terms in this table are the same as those in Table I of Ref. \cite{Gao:2019liu} except for the two new terms $\omega_3(\Gamma)$ and $\omega_3(\Gamma)K$. }
\lb{blocks}
\begin{ruledtabular}
\begin{tabular} {c|c|c}
$d$  & $(d_t, d_s)$ &  Operators\\
 \hline
 0 & $(0, 0)$ & 1 \\
\hline
\multirow{2}*{1} & (1,0)  & $K$ \\
\cline{2-3}
& (0, 1) & - \\
\hline
\multirow{3}*{2} &  (2, 0) & $K_{ij}$, $K^2$ \\
\cline{2-3}
& (1, 1) & - \\
\cline{2-3}
& (0, 2) & $R$ \\
\hline
\multirow{4}*{3} & (3, 0) & $K_{ij} K^{jk} K^i_k$, $K_{ij}K^{ij} K$, $K^3$ \\
\cline{2-3}
& (2, 1) & $\varepsilon_{ijk} K^i_l \nabla^j K^{kl}$ \\
\cline{2-3} 
& (1, 2) & $\nabla^i \nabla^j K_{ij}$, $\nabla^2 K$, $R^{ij} K_{ij}$, $RK$ \\
\cline{2-3}
& (0, 3) & $\omega_3(\Gamma)$ \\
\hline
\multirow{5}*{4} & (4, 0) & $K_{i j} K^{j k} K_{k}^{i} K$, $\left(K_{i j} K^{i j}\right)^{2}$, $K_{i j} K^{i j} K^{2}$,  $K^{4}$ \\
\cline{2-3}
& (3, 1) & $ \varepsilon_{i j k} \nabla_{m} K_{n}^{i} K^{j m} K^{k n}, \quad \varepsilon_{i j k} \nabla^{i} K_{m}^{j} K_{n}^{k} K^{m n}, \quad \varepsilon_{i j k} \nabla^{i} K_{l}^{j} K^{k l} K$ \\
\cline{2-3}
& (2, 2) & $\nabla_{k} K_{i j} \nabla^{k} K^{i j}$, $\nabla_{i} K_{j k} \nabla^{k} K^{i j}$, $\nabla_{i} K^{i j} \nabla_{k} K_{j}^{k}$, $\nabla_{i} K^{i j} \nabla_{j} K$, $\nabla_{i} K \nabla^{i} K$, $R_{i j} K_{k}^{i} K^{j k}$, $R K_{i j} K^{i j}$,  $R_{i j} K^{i j} K $, $RK^{2}$ \\
\cline{2-3}
& (1, 3) & $\varepsilon_{i j k} R^{i l} \nabla^{j} K_{l}^{k}, \quad \varepsilon_{i j k} \nabla^{i} R_{l}^{j} K^{k l}$, $\omega_3(\Gamma) K$ \\
\cline{2-3}
& (0, 4) & $\nabla^{i} \nabla^{j} R_{i j}, \quad \nabla^{2} R, \quad R_{i j} R^{i j}, \quad R^{2}$
\end{tabular}
\end{ruledtabular}
\end{table*}

In order to construct a concrete gravitational action with spatial covariance, one first needs to specify the building blocks which are invariant under the spatial diffeomorphisms. These building blocks consist of linear combinations of the extrinsic curvature $K_{ij}$, intrinsic curvature $R_{ij}$, and their spatial derivatives and derivatives of the spatial metric itself. Up to the fourth order in derivatives of spatial metric variables $g_{ij}$, we have the building blocks as shown in Table \ref{blocks} which are all scalars under the spatial diffeomorphisms. Then, the general action of the gravitational part of the spatial covariant gravities is given by \cite{Gao:2019liu}
\bqn
S_g &=& \int dt d^3 x \sqrt{g} N \Big({\cal L}^{(0)} + {\cal L}^{(1)} + {\cal L}^{(2)} + {\cal L}^{(3)} + {\cal L}^{(4)} \nb \\
&&~~~~~~~~~~~~~~~~~~ ~~~~~~~  + \tilde {\cal L}^{(3)} + \tilde {\cal L}^{(4)}\Big), \lb{oldmodel}
\eqn
where ${\cal L}^{(0)},\; {\cal L}^{(1)}, \;{\cal L}^{(2)}, \; {\cal L}^{(3)}$, and ${\cal L}^{(4)}$ are the parity-preserving terms, which are given by
\bqn
\mathcal{L}^{(0)} &=& c_1^{(0,0)}, \\
\mathcal{L}^{(1)} &=& c_1^{(1,0)} K, \\
\mathcal{L}^{(2)} &=& c_1^{(2, 0)} K_{ij} K^{ij} + c_2^{(2, 0)} K^2 + c_1^{(0, 2)} R, \\
\mathcal{L}^{(3)} &=& c_1^{(3, 0)} K_{ij} K^{jk} K^i_k+ c_2^{(3, 0)} K_{ij}K^{ij} K + c_3^{(3, 0)} K^3 \nb\\
&& + c_1^{(1,2)} \nabla^i \nabla^j K_{ij} + c_2^{(1,2)} \nabla^2 K + c_3^{(1,2)} R^{ij} K_{ij} \nb\\
&& + c_4^{(1,2)} RK, \\
\mathcal{L}^{(4)} &=& c_1^{(4,0)} K_{i j} K^{j k} K_{k}^{i} K + c_2^{(4,0)} \left(K_{i j} K^{i j}\right)^{2} \nb\\
&& + c_3^{(4,0)} K_{i j} K^{i j} K^{2}+ c_4^{(4,0)} K^4 \nb\\
&& + c_1^{(2,2)} \nabla_{k} K_{i j} \nabla^{k} K^{i j} + c_2^{(2,2)} \nabla_{i} K_{j k} \nabla^{k} K^{i j}  \nb\\
&& +c_3^{(2,2)} \nabla_{i} K^{i j} \nabla_{k} K_{j}^{k}+c_4^{(2,2)}\nabla_{i} K^{i j} \nabla_{j} K \nb\\
&&+c_5^{(2,2)}\nabla_{i} K \nabla^{i} K +c_6^{(2,2)} R_{i j} K_{k}^{i} K^{j k}\nb\\
&& +c_7^{(2,2)} R K_{i j} K^{i j} +c_8^{(2,2)} R_{i j} K^{i j} K +c_9^{(2,2)} RK^{2}\nb\\
&&+ c_1^{(0,4)} \nabla^{i} \nabla^{j} R_{i j} + c_2^{(0,4)} \nabla^{2} R  + c_3^{(0,4)} R_{i j} R^{i j} \nb\\
&&+ c_4^{(0,4)}R^2, \lb{4order}
\eqn
and $\tilde{\cal L}^{(3)}$ and $\tilde{\cal L}^{(4)}$ are parity-violating terms which are given by
\bqn
\tilde{\cal L}^{(3)} &=&  c_1^{(2,1)} \varepsilon_{ijk} K^i_l \nabla^j K^{kl} + c_1^{(0,3)} \omega_3(\Gamma),  \lb{L3}\\
\tilde{\cal L}^{(4)} &=& c_1^{(3,1)} \varepsilon_{i j k} \nabla_{m} K_{n}^{i} K^{j m} K^{k n} + c_2^{(3,1)}\varepsilon_{i j k} \nabla^{i} K_{m}^{j} K_{n}^{k} K^{m n} \nb\\
&& + c_3^{(3,1)} \varepsilon_{i j k} \nabla^{i} K_{l}^{j} K^{k l} K + c_1^{(1,3)} \varepsilon_{i j k} R^{i l} \nabla^{j} K_{l}^{k} \nb\\
&& + c_2^{(1,3)} \varepsilon_{i j k} \nabla^{i} R_{l}^{j} K^{k l}+ c_3^{(1,3)} \omega_3(\Gamma) K. \lb{L4}
\eqn
All of the coefficients like $c_i^{(d_t, d_s)}$ are functions of $t$ and $N$. Note that in Table \ref{blocks} and Eqs.~(\ref{L3}, \ref{L4}), we add the spatial Chern-Simons term $ \omega_3(\Gamma)$ and its coupling to $K$, which are absent in the original action in Ref.~\cite{Gao:2019liu}. It is interesting to note that the above action reduces to GR if one imposes
\bqn
c_{1}^{(2,0)} = c_1^{(0,2)}=- c_2^{(2, 0)}=\frac{M_{\rm Pl}^2}{2} 
\eqn
where all other coefficients $c_{i}^{(d_t, d_s)}$ are set to zero and $M_{\rm Pl}$ is the reduced Planck mass.

\section{GWs in spatial covariant gravities}
\renewcommand{\theequation}{3.\arabic{equation}} \setcounter{equation}{0}

In this section, we consider the propagation of GWs of spatial covariant gravities in a homogeneous and isotropic background. The spatial covariant gravities can have three degrees of freedom propagating in the theory, of which two are tensorial and one is of the scalar type. The extra scalar mode, which is absent in GR, is in general expected to be small compared to the two observed tensorial modes. For this reason, hereafter we only consider the two tensorial modes of GWs (the transverse and traceless modes). In the flat Friedmann-Robertson-Walker spacetime, GW is described by the tensor perturbations of the metric, i.e., 
\bqn
ds^2 = a^2(\tau) [-d\tau^2 + (\delta_{ij} + h_{ij})dx^i dx^j], \lb{metric_spatia}
\eqn
where $a(\tau)$ is the scale factor of the expanding Universe and hereafter we set $a_0=1$ as its present value. $\tau$ denotes the conformal time, which is related to the cosmic time $t$ by $dt =a d\tau$. $h_{ij}$ denotes the GWs, which we take to be transverse and traceless, $\partial^i h_{ij} =0 = h^i_i$. Then, the action of GWs up to the quadratic action can be written in the form \cite{Gao:2019liu}, 
 \begin{widetext}
 \bqn
 S^{(2)} &=&  \int dtd^3 x \frac{a^3}{2}\Big[ {\cal G}_0(t)\dot h_{ij} \dot h^{ij} + {\cal G}_1(t) \epsilon^{ijk}\dot h_{li} \frac{1}{a} \partial_j \dot h^{l}_k  - {\cal G}_2(t) \dot h_{ij} \frac{\Delta}{a^2} \dot h^{ij}  \nb\\
 &&~~~~~~~~~~ + {\cal W}_0(t) h_{ij} \frac{\Delta}{a^2} h^{ij} + {\cal W}_1(t) \epsilon^{ijk} h_{li} \frac{1}{a} \frac{\Delta}{a^2} \partial_j h^l_k  - {\cal W}_2(t) h_{ij} \frac{\Delta^2}{a^4} h^{ij}\Big],\nb\\
 \eqn
 where ${\cal G}_n$ and ${\cal W}_n$ are given by \cite{Gao:2019liu} \footnote{In ${\cal W}_1$ we add the contributions from the two new terms $\omega_3(\Gamma)$ and $\omega_3(\Gamma)K$.}
 \bqn
 {\cal G}_0 &=& \frac{1}{2} \Big[c_1^{(2,0)} + 3 (c_1^{(3,0)} + c_2^{(3, 0)}) H + 3(3 c_1^{(4,0)} + 2 c_2^{(4,0)} + 3 c_3^{(4,0)}) H^2 \Big], \\
 {\cal G}_1 &=& \frac{1}{2} \Big[c_1^{(2,1)}- (c_1^{(3,1)} - 2 c_2^{(3,1)} - 3 c_3^{(3,1)}) H \Big], \\
 {\cal G}_{2} &=& \frac{1}{2}c_1^{(2,2)}, \\
 {\cal W}_0 &=&\frac{1}{2}\Big[ c_1^{(0, 2)} + \frac{1}{2} \dot c_3^{(1,2)} + \frac{1}{2} \Big(3 c_3^{(1,2)} + 6 c_4^{(1,2)} + 2 \dot c_6^{(2,2)} + 3 \dot c_8^{(2,2)}\Big)  H \nb\\
 && + \frac{1}{2}\Big(4c_6^{(2,2)} + 6 c_7^{(2,2)} + 9 c_8^{(2,2)} + 18 c_9^{(2,2)}\Big) H^2 + \frac{1}{2} \Big(2c_6^{(2,2)} + 3 c_8^{(2,2)}\Big)\dot H \Big], \\
 {\cal W}_1 &=& \frac{1}{4} \Big(\dot c_1^{(1,3)} + \dot{c}_2^{(1,3)}\Big) + c_1^{(0,3)} - 3  c_{3}^{(1,3)} H, \\
 {\cal W}_2 &=& - \frac{1}{2}c_3^{(0,4)}.
 \eqn
  \end{widetext}
 Here a dot denotes a derivative with respect to the cosmic time $t$, $H=\dot a/a$ is the Hubble parameter, and $\Delta \equiv \delta^{ij}\partial_i \partial_j$ with $\delta^{ij}$ being the Kronecker delta. We consider the GWs propagating in  the homogeneous and isotropic
background , and ignore the source term. With the above action, one can obtain the equation of motion for $h_{ij}$ as
 \bqn
 && \left({\cal G}_0 - {\cal G}_2 \frac{\partial^2}{a^2}\right)h''_{ij} + \Big[2 \mathcal{H}{\cal G}_0 + {\cal G}_0' - {\cal G}_2' \frac{\partial^2}{a^2} \Big] h'_{ij} \nb\\
 &&- \left[ {\cal W}_0 - {\cal W}_2 \frac{\partial^2}{a^2} \right]\partial^2 h_{ij} \nb\\
 && + \epsilon_{ilk} \frac{\partial^l}{a} \left[{\cal G}_1 \partial_\tau^2 + (2 {\cal H} {\cal G}_1 + {\cal G}_1') \partial_\tau - {\cal W}_1 \partial^2\right] h_{j}^k =0,\nb\\
 \eqn
 where $\mathcal{H} \equiv a'/a$ and a prime denotes a derivative with respect to the conformal time $\tau$.
 
 In order to study the propagation of GWs in the spatial covariant gravities, it is convenient to decompose the GWs into the circular polarization modes. To study the evolution of $h_{ij}$, we expand it over spatial Fourier harmonics,
 \bqn
 h_{ij}(\tau, x^i) = \sum_{A={\rm R, L}} \int \frac{d^3k}{(2\pi)^3} h_A(\tau, k^i) e^{i k_i x^i} e_{ij}^A(k^i),\nb\\
 \eqn
 where $e_{ij}^A$ denotes the circular polarization tensors and satisfies the relation
 \bqn
 \epsilon^{ijk} n_i e_{kl}^A = i \rho_A e^{j A}_l,
 \eqn
 with $\rho_{\rm R} =1$ and $\rho_{\rm L} = -1$. We find that the propagation equations of these two modes are decoupled, which can be cast in the form \cite{Gao:2019liu}
 \bqn\lb{eom_A}
 h''_A + (2+\Gamma_A) \mathcal{H} h'_A + \omega_A^2 h_A=0,
 \eqn
 where
 \bqn
 {\cal H}\Gamma_A &=& \left[\ln \left({\cal G}_0 +  \rho_A {\cal G}_1 \frac{k}{a} + {\cal G}_2 \frac{k^2}{a^2}\right) \right]',\\
 \frac{\omega_A^2}{k^2} &=& \frac{{\cal W}_0 +  \rho_A {\cal W}_1 \frac{k}{a} + {\cal W}_2 \frac{k^2}{a^2}}{{\cal G}_0 +  \rho_A {\cal G}_1 \frac{k}{a} + {\cal G}_2 \frac{k^2}{a^2}}.
 \eqn
 The properties of the propagation of GWs with nonzero $\Gamma_A$ and a modified dispersion relation $\omega_k^2$ were discussed in Ref. \cite{Gao:2019liu}. Several specific forms of the spatial covariant gravities in which the GWs propagate at the speed of light were also explored \cite{Gao:2019liu}. The derivations of the spatial covariant gravities from GR are fully characterized by the quantities $\Gamma_A$ and $\omega_A^2$. The former represents the corrections to the damping rate which modifies the amplitude damping rate of the GWs during their propagation in the cosmological background, and the latter is the modified dispersion relation of GWs which leads to a phase shifting of GWs from distant sources.  
 
We expect that the derivations from GR are small, such that
\bqn
\Gamma_A \ll 1, \;\; \left|\frac{\omega_A^2}{k^2} -1\right| \ll 1.
\eqn
Thus, we can consider all of the new effects on GWs beyond GR as small corrections to the standard GR result. In this way, we are able to expand ${\cal H}\Gamma_A$ and $\omega_A$ as
\bqn
{\cal H}\Gamma_A &\simeq& (\ln {\cal G}_0)' + \frac{\rho_A}{{\cal G}_0} \left({\cal G}_1 \frac{k}{a}\right)' + \frac{1}{{\cal G}_0}\left({\cal G}_2 \frac{k^2}{a^2}\right)', \lb{gamma}\\
\frac{\omega_A^2}{k^2}& \simeq& \frac{{\cal W}_0}{{\cal G}_0} +  \rho_A \frac{{\cal W}_1 - {\cal G}_1}{{\cal G}_0} \frac{k}{a} + \frac{{\cal W}_2 - {\cal G}_2}{{\cal G}_0} \frac{k^2}{a^2}.\nb \lb{omega}\\
\eqn
Note that in the above expansion we only consider the first-order terms of each coefficient, i.e., $1-{\cal W}_0/{\cal G}_0$, ${\cal W}_1$, ${\cal G}_1$, ${\cal W}_2$, and ${\cal G}_2$. 

With these considerations, the equation of motion (\ref{eom_A}) can be further simplified into the standard parametrized form \cite{Zhao:2019xmm}
\bqn
h_A'' + (2 + \bar \nu +\nu_A) {\cal H} h_A' + (1+\bar \mu + \mu_A) k^2 h_A=0,\nb \lb{parameter_eom}\\
\eqn
with
\bqn
{\cal H} \bar \nu &=&  (\ln {\cal G}_0)' + \frac{1}{{\cal G}_0}\left({\cal G}_2 \frac{k^2}{a^2}\right)', \lb{GammaA}\\
{\cal H} \nu_A &=&   \frac{\rho_A }{{\cal G}_0}\left({\cal G}_1 \frac{k}{a}\right)', \\
\bar \mu &=&  \frac{{\cal W}_0}{{\cal G}_0}-1 + \frac{{\cal W}_2 - {\cal G}_2}{{\cal G}_0} \frac{k^2}{a^2},\\
\mu_A &=&   \rho_A \frac{{\cal W}_1 - {\cal G}_1}{{\cal G}_0} \frac{k}{a}.
\eqn
In such a parametrization, the new effects arising from theories beyond GR are characterized by four parameters: $\bar \nu$, $\bar \mu$, $\nu_A$ and $\mu_A$. The parameters $\nu_A$, and $\mu_A$ label the effects of the parity-violating terms in the spatial covariant gravities, and $\bar \nu$ and $\bar \mu$ describe the effects of other possible modifications that are not relevant to parity violation. Among these four parameters, $\mu_A$ and $\bar \mu$ determine the speed of GWs, while $\nu_A$ and $\bar \nu$ determine the damping rate of GWs during their propagation.  

The coefficients $1-{\cal W}_0/{\cal G}_0$, ${\cal W}_1$, ${\cal G}_1$, ${\cal W}_2$, and ${\cal G}_2$ arising from different terms in the spatial covariant gravity induce distinct effects on GW propagation. In order to study the effects of the spatial covariant gravity on GW propagation, we separately consider each term in Eqs. (\ref{gamma}) and (\ref{omega}) by setting the others to zero. 
In this way, the four parameters $\bar \nu$, $\bar \mu$, $\nu_A$, and $\mu_A$ can be further parametrized in the following form \cite{Zhao:2019xmm}:
\bqn
\mathcal{H} \bar{\nu} &=&\left[\alpha_{\bar{\nu}}(\tau)\left(k / a M_{\mathrm{LV}}\right)^{\beta_{\bar{\nu}}}\right]', \\
 \bar{\mu}&=&\alpha_{\bar{\mu}}(\tau)\left(k / a M_{\mathrm{LV}}\right)^{\beta_{\bar{\mu}}}, \\
\mathcal{H} \nu_{\mathrm{A}} &=&\left[\rho_{\mathrm{A}} \alpha_{\nu}(\tau)\left(k / a M_{\mathrm{PV}}\right)^{\beta_{\nu}}\right]^{\prime}, \\
\mu_{\mathrm{A}}&=&\rho_{\mathrm{A}} \alpha_{\mu}(\tau)\left(k / a M_{\mathrm{PV}}\right)^{\beta_{\mu}},
\eqn
where $\beta_{\bar \nu}$ and $\beta_{\bar \mu}$ are arbitrary even numbers and $\beta_\nu$ and $\beta_\mu$ are arbitrary odd numbers. $\alpha_{\bar \nu}$, $\alpha_{\bar \mu}$, $\alpha_\nu$, and $\alpha_\mu$ are the arbitrary functions of time. To write this form, we separately consider each term in Eqs. (\ref{gamma}) and (\ref{omega}) and set the others to zero. The different terms in Eqs. (\ref{gamma}) and (\ref{omega}) correspond to different values of the above parameters. The corresponding values of the parameters $(\alpha_{\bar \nu}, \beta_{\bar \nu}, \alpha_{\bar \mu}, \beta_{\bar \mu}, \alpha_{\nu}, \beta_\nu, \alpha_\mu, \beta_\mu)$ for the different terms defined in Eqs. (\ref{gamma}) and (\ref{omega}) are listed in Table \ref{paramaters}. In Table \ref{specific_theories} we present the corresponding values of the parameters $(\alpha_{\bar \nu}, \beta_{\bar \nu}, \alpha_{\bar \mu}, \beta_{\bar \mu}, \alpha_{\nu}, \beta_\nu, \alpha_\mu, \beta_\mu)$ for several specific scalar-tensor theories that can be related to spatial covariant gravities in the unitary gauge. 

\begin{table*}
\caption{\label{paramaters}%
Corresponding values of the parameters $(\alpha_{\bar \nu}, \beta_{\bar \nu}, \alpha_{\bar \mu}, \beta_{\bar \mu}, \alpha_{\nu}, \beta_\nu, \alpha_\mu, \beta_\mu)$ for the different terms in defined in Eqs. (\ref{gamma}) and (\ref{omega}). }
\begin{ruledtabular}
\begin{tabular}{l|ccccccccl}
&\multicolumn{2}{c}{${\cal H} \bar \nu$} &\multicolumn{2}{c}{${\bar \mu}$} & \multicolumn{2}{c}{${\cal H}{\nu_A}$} & \multicolumn{2}{c}{${\mu_A}$}\\
\cline{2-3}  \cline{4-5} \cline{6-7} \cline{8-9}
  & $\alpha_{\bar \nu}$ & $\beta_{\bar \nu}$ & $\alpha_{\bar \mu}$ & $\beta_{\bar \mu}$ & $\alpha_{\nu}$ & $\beta_{\nu}$ & $\alpha_{\mu}$ & $\beta_{\mu}$  & Related coefficients \\
  \colrule
     $ {\cal G}_0$ &    $\ln{\cal G}_0$ & 0 & $-1+{\cal W}_0/{\cal G}_0$ & 0 & --- & --- & ---& --- & $c_1^{(2,0)}, c_1^{(3,0)}, c_2^{(3,0)}, c_1^{(4,0)}, c_2^{(4,0)}, c_3^{(4,0)}$ \\
     ${\cal G}_1$ & --- & --- & ---& --- & ${\cal G}_1 M_{\rm PV}/{\cal G}_0$ & 1 & $- {\cal G}_1 M_{\rm PV}/{\cal G}_0$ & 1 &  $c_1^{(2,1)}, c_1^{(3,1)}, c_2^{(3,1)}, c_3^{(3,1)}$ \\
     ${\cal G}_2$ & ${\cal G}_2 M_{\rm LV}^2/{\cal G}_0$   &  2 & $- {\cal G}_2 M_{\rm LV}^2/{\cal G}_0$& 2& --- & --- & ---& --- & $c_1^{(2,2)}$ \\
     ${\cal W}_0$ & --- & ---  & $-1+{\cal W}_0/{\cal G}_0$ & 0 &  --- & --- & ---& --- & $c_1^{(0,2)}, c_3^{(1,2)}, c_4^{(1,2)}, c_6^{(2,2)}, c_7^{(2,2)},c_8^{(2,2)}, c_9^{(2,2)}$  \\
     ${\cal W}_1$  & --- & --- & ---& --- &   --- & --- &  ${\cal W}_1 M_{\rm PV}/{\cal G}_0$& 1 & $ c_1^{(0,3)}, c_1^{(1,3)}, c_2^{(1,3)}, c_3^{(1,3)} $ \\
     ${\cal W}_2$& --- & --- & ${\cal G}_2 M_{\rm LV}^2/{\cal G}_0$& 2&   --- & --- & --- &  --- & $c_3^{(0,4)}$\\
\end{tabular}
\end{ruledtabular}
\end{table*}

\begin{table*}
\caption{\label{specific_theories}%
Corresponding values of the parameters $(\alpha_{\bar \nu}, \beta_{\bar \nu}, \alpha_{\bar \mu}, \beta_{\bar \mu}, \alpha_{\nu}, \beta_\nu, \alpha_\mu, \beta_\mu)$ for several specific scalar-tensor theories that can be related to spatial covariant gravities in the unitary gauge. }
\begin{ruledtabular}
\begin{tabular}{l|cccccccl}
&\multicolumn{2}{c}{${\cal H} \bar \nu$} &\multicolumn{2}{c}{${\bar \mu}$} & \multicolumn{2}{c}{${\cal H}{\nu_A}$} & \multicolumn{2}{c}{${\mu_A}$}\\
\cline{2-3}  \cline{4-5} \cline{6-7} \cline{8-9}
  & $\alpha_{\bar \nu}$ & $\beta_{\bar \nu}$ & $\alpha_{\bar \mu}$ & $\beta_{\bar \mu}$ & $\alpha_{\nu}$ & $\beta_{\nu}$ & $\alpha_{\mu}$ & $\beta_{\mu}$   \\
  \colrule
     Horndeski &   $\ln\frac{b_0-3c_0 H}{2}$ & 0 & $\sqrt{\frac{d_1-\dot a_1}{b_0-3c_0 H}}-1$ & 0 & --- & --- & ---& --- \\
     Scalar-Gauss-Bonnet & $\ln\frac{1+8\dot{\xi} H}{M_{\rm Pl}^2}$ & 0 & $8 \frac{\ddot{\xi} H + \dot{\xi} \dot{H}}{M_{\rm Pl}^2 H}$& 0 & ---& --- &--- & --- \\
    Chern-Simons & ---  &  --- & --- & ---&  $- \dot \vartheta M_{\rm PV}$& 1 & ---& --- \\
     Lorentz-violating Weyl & --- & ---  & $4 \gamma M_{\rm LV}^2$ & 2 &  --- & --- & ---& ---   \\
    Chiral scalar-tensor  & --- & --- & ---& --- &   \makecell{$M_{\rm PV} \Big(- \dot \vartheta + 2(a_3+2a_1)\dot \phi^2 H $ \\ $-2b_1 \dot \phi^3 + 2(b_4+b_5-b_3) \dot \phi^4 H \Big)$} & 1 &  \makecell{$M_{\rm PV} \Big(2a^2 \partial_t\big[(2a_1+a_3) \dot \dot \phi^2 a^{-2}\big] $ \\ $-2b_1 \dot \phi^3 + 2(b_4+b_5-b_3) \dot \phi^4 H \Big)$}  & 1  \\
\end{tabular}
\end{ruledtabular}
\end{table*}

\section{Effects of  the spatial covariant gravities on GWs and their constraints }
\renewcommand{\theequation}{4.\arabic{equation}} \setcounter{equation}{0}

\subsection{Frequency-independent effects from ${\cal G}_0$ and ${\cal W}_0$}

The coefficients ${\cal G}_0$ and ${\cal W}_0$ induce two distinct and frequency-independent effects on the propagation of GWs. One is the modification of the speed of the GWs if ${\cal G}_0 \neq {\cal W}_0$, and another is the modified fraction term of the GWs if $(\ln{\cal G}_0)'$ is nonzero.  In the following subsubsections, we discuss these individually. 

\subsubsection{Modification of speed of GWs}

When ${\cal W}_0 \neq {\cal G}_0$, the speed of the GWs is modified in a frequency-independent manner,
\bqn
c_{\rm gw} = \sqrt{\frac{{\cal W}_0}{{\cal G}_0}}.
\eqn
For a GW event with an electromagnetic counterpart, $c_{\rm gw}$ can be constrained by comparison with the arrival time of the photons. For the binary neutron star merger GW170817 and its associated electromagnetic counterpart GRB170817A, the almost coincident observation of both the electromagnetic wave and GW places an exquisite bound on the GW speed \cite{LIGOScientific:2017vwq, LIGOScientific:2017zic},
\bqn
-3 \times 10^{-15} \leq c_{\rm gw} -1 \leq 7\times 10^{-16}.
\eqn
Note that here we set the speed of light $c=1$. This bound then leads to a constraint on ${\cal W}_0/{\cal G}_0$ as
\bqn
-3 \times 10^{-15} \leq \sqrt{\frac{{\cal W}_0}{{\cal G}_0}}-1  \leq 7\times 10^{-16}.
\eqn
From this constraint, one has
\begin{widetext}
\bqn
&&-3 \times 10^{-15} \leq \frac{1}{M_{\rm Pl}^{2}}\Bigg[\delta c_1^{(0, 2)} - \delta c_1^{(2,0)}+ \frac{1}{2} \dot c_3^{(1,2)} -3 (c_1^{(3,0)} + c_2^{(3, 0)}) H+  \Big(\frac{3}{2} c_3^{(1,2)} + 3 c_4^{(1,2)} +  \dot c_6^{(2,2)} + \frac{3}{2} \dot c_8^{(2,2)}\Big)  H \nb\\
 && + \Big(2c_6^{(2,2)} + 3 c_7^{(2,2)} + \frac{9}{2} c_8^{(2,2)} + 9 c_9^{(2,2)}\Big) H^2 - (9 c_1^{(4,0)} + 6 c_2^{(4,0)} + 9 c_3^{(4,0)}) H^2+ \Big(c_6^{(2,2)} + \frac{3}{2}  c_8^{(2,2)}\Big)\dot H \Bigg] \leq 7\times 10^{-16}.\nb\\
 \eqn
\end{widetext}
Here $\delta c_1^{(0,2)} \equiv c_1^{(0,2)} - \frac{1}{2} M_{\rm Pl}^2$ and $\delta c_1^{(2,0)} \equiv c_1^{(2,0)}- \frac{1}{2} M_{\rm Pl}^2$. In deriving the above bound, we have expanded all of the coefficients $c_{i}^{(d_t, d_s)}$ beyond GR in the modified speed of GWs to the first order.

\subsubsection{Modified GW friction from $\ln{\cal G}_0$}

The term $(\ln{\cal G}_0)'$ in Eq.~(\ref{GammaA}) also induces an additional friction term in the propagation equation of GWs. In GR, ${\cal G}_0$ is related to the Planck mass $M_{\rm Pl}^2$ through ${\cal G}_0 = M_{\rm Pl}^2/4$ and thus $(\ln{\cal G}_0)'=0$. In the spatial covariant gravities, ${\cal G}_0$ is time dependent and one can introduce an effective and time-dependent Planck mass $M_{*}(t)$ by writing ${\cal G}_0 = M_{*}^2(t)/4$. Then the modified friction term $(\ln{\cal G}_0)'$ can be written in terms of the running of the effective Planck mass in the form
\bqn
(\ln{\cal G}_0)' = H \frac{d\ln M_*^2}{\ln a}.
\eqn
Such an additional friction term also changes the damping rate of GWs during propagation. This leads to a GW luminosity distance $d_L^{\rm gw}$ which is related to the standard luminosity distance of electromagnetic signals $d_L^{\rm em}$ as \cite{Belgacem:2018lbp, Belgacem:2017ihm}
\bqn
d_L^{\rm gw}(z) &=& d_L^{\rm em}(z) \exp \Big\{\frac{1}{2} \int_0^{z} \frac{dz'}{1+z'} \frac{(\ln{\cal G}_0)'}{\cal H}\Big\} \nb\\
&=& d_L^{\rm em}(z) \exp \Big\{ \frac{1}{2} \int  d (\ln{\cal G}_0) \Big\}.
\eqn
Thus, it is possible to probe GW friction $(\ln {\cal G}_0)'$ using the multimessenger measurements of $d_L^{\rm gw}$ and $d_L^{\rm em}$.   

However, such a probe relies sensitively on the time evolution of $(\ln {\cal G}_0)'$, which is in general unknown. In order to probe GW friction, there are two approaches to parametrize the time evolution of $(\ln {\cal G}_0)'$. One is $c_M$-parametrization \cite{Lagos:2019kds}, which is based on the evolution of the dark energy in the Universe, and another is $\Xi$-parametrization \cite{Belgacem:2018lbp}, which is a theory-based parametrization that can fit a lot of modified gravities. 

For $c_M$-parametrization, the GW friction is written as \cite{Lagos:2019kds}
\bqn
\frac{(\ln{\cal G}_0)'}{\cal H} = c_M \frac{\Omega_\Lambda(z)}{\Omega_\Lambda(0)},
\eqn
where $z$ is the redshift of the GW source and $\Omega_\Lambda$ is the fractional dark energy density. If one considers the dark energy density as a constant, then one has \cite{Leyde:2022orh}
\bqn
\Omega_\Lambda(z) = \frac{\Omega_\Lambda(0)}{\Omega_\Lambda(0)+ (1+z)^3 \Omega_m(0)},
\eqn
where $\Omega_m(0)$ is the value of the fractional energy density of matter. Several constraints on $c_M$ have been derived using both $d_L^{\rm gw}$ and $d_L^{\rm em}$ from GW events or populations \cite{Leyde:2022orh, Mastrogiovanni:2020mvm, Ezquiaga:2021ayr}. Here we adopt a constraint from Ref. \cite{Leyde:2022orh} from a jointed parameter estimation of the mass distribution, redshift evolution, and GW friction with GWTC-3 for different BBH population models, which gives
\bqn
c_M = -0.6^{+2.2}_{-1.2}.
\eqn
This corresponds to
\bqn
\left.\frac{(\ln{\cal G}_0)'}{\cal H} \right|_{z=0} =  -0.6^{+2.2}_{-1.2}.
\eqn
 
For $\Xi$ parametrization, the full redshift dependence of GW friction is described by two parameters $(\Xi_0, n)$, with which the ratio between the GW and electromagnetic luminosity distances can be written as \cite{Belgacem:2018lbp}
 \bqn
 \frac{d_L^{\rm gw}(z)}{d_L^{\rm em} (z)} \equiv \Xi(z) = \Xi_0 + \frac{1-\Xi_0}{(1+z)^n}.
 \eqn
Such a parametrization corresponds to
 \bqn
 \frac{(\ln{\cal G}_0)'}{\cal H}  = \frac{2 n (1- \Xi_0)}{1-\Xi_0+\Xi_0(1+z)^n}.
 \eqn
 The relation between the $\Xi$-parametrization and $c_M$-parametrization was explored in Ref. \cite{Mancarella:2021ecn}. Several constraints on $(\Xi, n)$ were obtained using GW events with redshift information inferred from the corresponding electromagnetic counterparts \cite{Mastrogiovanni:2020mvm} or host galaxies \cite{Ezquiaga:2021ayr}, or the binary black hole mass function \cite{Mancarella:2021ecn}. A recent constraint on $(\Xi_0, n)$ was derived from an analysis of GW data in GWTC-3 with a BBH mass function, which gives \cite{Mancarella:2021ecn}
 \bqn
 \Xi_0 = 1.0^{+0.6}_{-0.5}, \;\; n=2.5^{+1.7}_{-1.1}
 \eqn
 with a prior uniform in $\ln \Xi_0$. This bound leads to a constraint on $(\ln {\cal G})'$ in the form
 \bqn
 -3.0 < \left. \frac{(\ln{\cal G}_0)'}{\cal H} \right|_{z=0} < 2.5.
 \eqn
This leads to
\begin{widetext}
\bqn
-3.0 < \frac{1}{H} \partial_t \ln  \Big[c_1^{(2,0)} + 3 (c_1^{(3,0)} + c_2^{(3, 0)}) H + 3(3 c_1^{(4,0)} + 2 c_2^{(4,0)} + 3 c_3^{(4,0)}) H^2 \Big] < 2.5.
\eqn
\end{widetext}

 \subsection{Parity-violating effects from ${\cal W}_1$ and ${\cal G}_1$}
 
 Adding the parity-violating terms introduced in Eqs. (\ref{L3}) and (\ref{L4}) to the action of the spatial covariant gravities leads to nonzero coefficients ${\cal W}_1$ and ${\cal G}_1$. Due to the parity violation, these two coefficients induce two distinct birefringent effects on the propagation of GWs: the amplitude birefringences and the velocity birefringences.
 
 \subsubsection{Amplitude birefringences of GWs from $({\cal G}_1 k/a)'$}
 
 The effects of the coefficient $({\cal G}_1 k/a)'$ are fully characterized by the parameter $\nu_A$, which leads to different damping rates for the left- and right-hand circular polarizations of GWs, so the amplitude of the left-hand circular polarization of GWs will increase (or decrease) during the propagation, while the amplitude of the right-hand modes will decrease (or increase). This effect induces modifications in the amplitude of the GW waveform in the form \cite{Zhao:2019xmm}
 \bqn
 h_A = h^{\rm GR}_A \exp{\left(-\frac{1}{2}\int_{\tau_e}^{\tau_0} {\cal H} \nu_A\right) d\tau} = h^{\rm GR}_A e^{\rho_A \delta h_1}, \nb\\
 \eqn
 with
 \bqn
 \delta h_1 &=& - \frac{1}{2} \left.\left[a_\nu \left(\frac{k}{aM_{\rm PV}}\right)^{\beta_\nu}\right]\right|^{a_0}_{a_e} = - \left.\frac{1}{2} \frac{{\cal G}_1 k}{{\cal G}_0 a} \right|^{a_0}_{a_e},\nb\\
 \eqn
 where $h^{\rm GR}$ denotes the waveform of GWs in GR, $a_0=a(t_0)$ with $t_0$ denoting the arrival time of GWs and $a_e = a(t_e)$ with $t_e$ being the emitted time.  We can convert the left- and right-hand GW polarization modes into the plus and cross modes which are used more often in GW detections. Using the relation
 \bqn
 h_{+} = \frac{h_{L} + h_{R}}{\sqrt{2}}, \;\; h_{\times} = \frac{h_L - h_R}{\sqrt{2} i},
 \eqn
 we obtain
 \bqn
 h_{+}(f) = h_{+}^{\rm GR} \cosh(\delta h_1) - i h_{\times}^{\rm GR}\sinh(\delta h_1), \\
 h_{\times}(f) = h_{\times}^{\rm GR} \cosh(\delta h_1) + i h_{+}^{\rm GR} \sinh(\delta h_1).
 \eqn
 
With the modified waveform in the above, one is able to test the amplitude birefringent effect induced by $({\cal G}_1 k/a)'/{\cal G}_0$  by comparing the modified waveform with the GW strain data from the GW detectors. By performing a Bayesian parameter estimation on the 12 LIGO-Virgo O1/O2 events with the modified waveform, the amplitude birefringent effect including parity violation has been constrained, which leads to a combined lower bound on the corresponding energy scale $M_{\rm PV}$ of \cite{Wang:2020cub, Yamada:2020zvt}
\bqn
M_{\rm PV} \gtrsim 10^{-22}\; {\rm GeV},
\eqn
which is a rather loose result. This is because GW detection is less sensitive to amplitude modification than phase. 

\begin{figure}
{
\includegraphics[width=8.1cm]{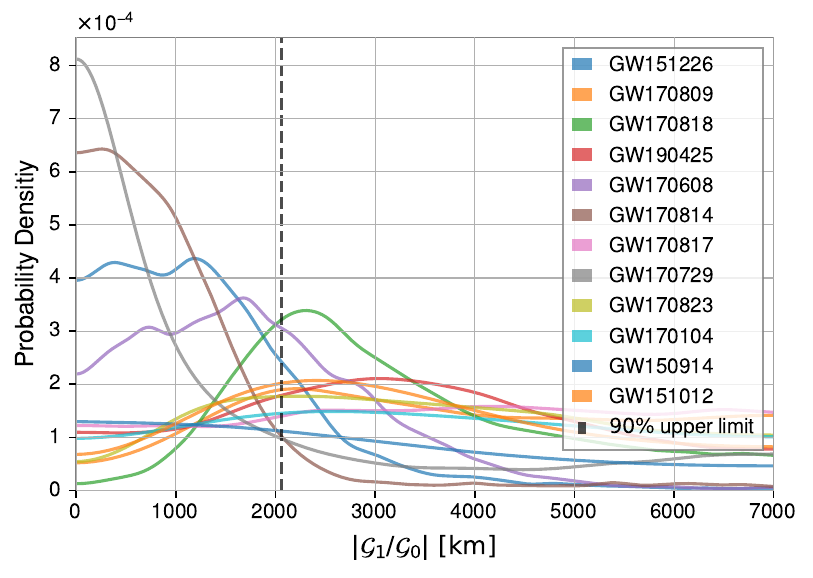}}\\
\caption{The posterior distributions for $|{\cal G}_1|$ from twelve LIGO-Virgo O1/O2 GW events. The vertical dash line denotes the 90\% upper limits of $|{\cal G}_1|$ from the combined result. } \label{av_combine}
\end{figure}

Here we would like to derive the constraint on the coefficient ${\cal G}_1$ by directly using the posterior samples obtained in Ref.\cite{Wang:2020cub} to test the amplitude birefringent effect with  12 LIGO-Virgo  O1/O2 events. The data for all 12 GW events are available in Ref. \cite{PV_web1}. In Ref. \cite{Wang:2020cub}, the amplitude birefringent effect due to parity violation was described by a parameter $A_\nu$, which can be related to ${\cal G}_1$ in the spatial covariant gravity via
\bqn
\delta h_1 = - A_\nu \pi f = - \left.\frac{1}{2} \frac{{\cal G}_1 k}{{\cal G}_0 a}\right|^{a_0}_{a_e}.
\eqn 
Thus, one has
\bqn
 A_\nu = \left.\frac{{\cal G}_1}{{\cal G}_0}\right|_{z=0} - \frac{{\cal G}_1(z)}{{\cal G}_0(z)} (1+z).
\eqn
Here $z$ is the redshift of the GW source. In principle, the coefficient ${\cal G}_1$ is an arbitrary function of time which can only be determined given a specific model of spatial covariant gravity. Considering that the redshifts of all 12 GW sources are not large, we can approximately treat ${\cal G}_1$ as constant, i.e., ignore its time dependence. Then, one can relate $A_\nu$ to ${\cal G}_1$ by
\bqn
\frac{{\cal G}_1}{{\cal G}_0} = -\frac{A_\nu}{z}.
\eqn
Then, from the posterior distributions of $A_\nu$ and the redshift $z$ obtained in Ref. \cite{Wang:2020cub} for each GW event, one can calculate the posterior distribution of ${\cal G}_1$ for each GW event. We plot the posterior probability distributions of $|{\cal G}_1|$ in Fig.~\ref{av_combine}. From this figure, we find that the posterior probability distributions  of ${\cal G}_1$ with 90\% confidence intervals are consistent with the GR value ${\cal G}_1=0$ for all 12 GW events.

In the above analysis, we have treated the quantity ${\cal G}_1$ as a constant. In this sense, this quantity is also a universal quantity for all GW events. Thus, one can combine all 12 individual posteriors of ${\cal G}_1$ to get the overall constraint. This can be done by multiplying the posterior distributions of the 12 GW events in LOGO-Virgo O1/O2 and then we find that the coefficient ${\cal G}_1$ can be constrained to be
\bqn
|{\cal G}_1/{\cal G}_0| < 2065 \; {\rm km} 
\eqn
at the 90\% confidence level. This constraint can be converted into the constraint on the combination of coefficients $c_1^{(2,1)}$, $c_1^{(3,1)}$, $c_2^{(3,1)}$, and $c_3^{(3,1)}$ as 
\bqn
\frac{|c_1^{(2,1)}- (c_1^{(3,1)} - 2 c_2^{(3,1)} - 3 c_3^{(3,1)}) H|}{M_{\rm Pl}^2}  < 1033 \;{\rm km}. \lb{av_constraint1}\nb\\ 
\eqn

The above results are obtained from the Bayesian parameter estimation in Ref. \cite{Wang:2020cub} by comparing the modified waveform with the GW strain data of the 12 GW events in the LIGO-Virgo O1/O2 catalog. Here we would like to mention that the amplitude birefringence also affects the statistical distribution of $\cos{\iota}$ over the population of binary black hole mergers \cite{Okounkova:2021xjv} with $\iota$ being the inclination angle of the binary black hole system. One can consider that the Universe is homogeneous and isotropic on cosmological scales, and that gravitational physics does not have any preferred direction. This implies that the underlying distribution for $\cos{\iota}$ is flat, meaning that its distribution is symmetric about zero when the amplitude birefringence is absent. When the amplitude birefringence induced by ${\cal G}_1$ is included, the distribution of $\cos\iota$ will preferentially have  $\cos\iota >0 (<0)$  if ${\cal G}_1 <0 (>0)$ \cite{Okounkova:2021xjv}. By checking the posterior distribution of $\cos\iota$ for GW events in the GWTC-2 catalog, we can impose a constraint on ${\cal G}_1$ \cite{Okounkova:2021xjv}, i.e., 
\bqn
|{\cal G}_1/{\cal G}_0| \lesssim 1000\; {\rm km}.
\eqn
This bound corresponds to 
\bqn
M_{\rm Pl}^{-2}\left|c_1^{(2,1)}- (c_1^{(3,1)} - 2 c_2^{(3,1)} - 3 c_3^{(3,1)}) H \right| < 500\; {\rm km}. \nb\\
\eqn
Note that this constraint improves on that in Eq. (\ref{av_constraint1}) by a factor of 2. 

It is worth mentioning here that in some specific models, the coefficient ${\cal G}_1$ could oscillate periodically. One example is the axion-Chern-Simons theory studied in Refs. \cite{Jung:2020aem, axion1, axion2}. In this scenario, the axion oscillation can induce parametric resonance in GWs at a certain frequency, which can produce resonance peaks in GW signals. Searching for these peaks in GW signals can thus place stringent constraints on both the axion-gravity coupling and axion mass, which can in principle place a more stringent constraint on ${\cal G}_1$, depending sensitively on the specific coupling form of ${\cal G}_1$. 
 
 \subsubsection{Velocity birefringence of GWs from ${\cal W}_1 - {\cal G}_1$}
 
When ${\cal W}_1 \neq {\cal G}_1$,  the coefficient ${\cal W}_1 - {\cal G}_1$ induces a nonzero parameter $\mu_A = \rho_A ({\cal W}_1 - {\cal G}_1)k/({\cal G}_0 a)$, which determines the speed of the GWs. In particular, due to parity violation, the parameter $\mu_A$ [or, equivalently, $\rho_A  ({\cal W}_1 - {\cal G}_1)/{\cal G}_0$] has the opposite signs for left- and right-hand circular polarizations of GWs. This leads to different velocities for left- and right-hand circular polarizations of GWs, and therefore the arrival times of the two circular polarization modes could be different. This phenomenon is known as velocity birefringence. 

As shown in Ref. \cite{Zhao:2019xmm}, with velocity birefringence, the different circular polarization modes will have different phase velocities
\bqn
v_A &\simeq& 1- \frac{1}{2} \rho_A \alpha_\mu(\tau) \left(\frac{k}{aM_{\rm PV}}\right) \nb\\
&=& 1 - \frac{1}{2} \rho_A \frac{{\cal W}_1 - {\cal G}_1}{{\cal G}_0} \frac{k}{a}.
\eqn
Consider GWs emitted at two different times $t_e$ and $t_e'$, with wave number $k$ and $k'$, and  received at corresponding arrival times $t_0$ and $t_0'$;  then, the different velocities of different circular polarization modes lead to a difference of their arrival times,
\bqn
\Delta t_0 &=& (1+z) \Delta t_e + \frac{\rho_A}{2} \left(\frac{k^{\beta_\mu}}{M_{\rm PV}^{\beta_\mu}} -\frac{k‘^{\beta_\mu}}{M_{\rm PV}^{\beta_\mu}} \right)\int_{t_e}^{t_0} \frac{\alpha_\mu}{a^{\beta_\mu+1}} dt\nb\\
&=&(1+z)\Delta t_e + \frac{\rho_A \left(k-k' \right)}{2} \int_{t_e}^{t_0} \frac{{\cal W}_1 - {\cal G}_1}{{\cal G}_0a^{2}}dt.
\eqn
Here $\Delta t_e = t_e -t_e'$. This velocity difference induces a modification of the phase of the GW signal emitted from a binary compact star system. The modified GW waveform in the Fourier domain reads
\bqn
h_A(f) = h_A^{\rm GR} e^{i \rho_A \delta \Psi_1}, \lb{gw_velocity}
\eqn
where the phase correction $\Delta \Psi_1$ induced by the velocity difference is expressed as
\bqn
\delta \Psi_1 = A_\mu (\pi f)^2
\eqn
with
\bqn
A_\mu &=& \int_{t_e}^{t_0} \frac{{\cal W}_1 -{\cal G}_1}{{\cal G}_0a^2} dt \nb\\
&=& \frac{1}{H_0} \int_0^z \frac{({\cal W}_1 - {\cal G}_1)(1+z')dz'}{{\cal G}_0\sqrt{\Omega_m(1+z')^3+\Omega_\Lambda}}. \lb{Amu}
\eqn
Here we adopt $H_0 = 67.8\; {\rm km/s/Mpc}$, $\Omega_m = 0.308$, and $\Omega_\Lambda=0.692$.  

There are several different ways to test the velocity birefringence induced by $({\cal W}_1 - {\cal G}_1)/{\cal G}_0$ in spatial covariant gravity. In Ref. \cite{Nishizawa:2018srh}, the velocity birefringence was constrained by comparing the arrival times of GW170817 and GRB170817a, which gives
\bqn
|({\cal W}_1 - {\cal G}_1)/{\cal G}_0| \lesssim 10^{-11} \; {\rm km}. \lb{velocity_PV1}
\eqn
With the arrival time difference of left- and right-hand GWs induced by velocity birefringence, Ref. \cite{Zhao:2019szi} proposed a new method for constraining velocity birefringence in a model-independent way by measuring the difference in arrival times of two GW polarizations. With this method, it is expect to constrain $|({\cal W}_1 -{\cal G}_1)/{\cal G}_0|$ to be $\lesssim 10^{-14} \; {\rm km}$. This constraint is better than Eq. (\ref{velocity_PV1}) by 3 orders of magnitude. The velocity birefringence could also slightly widen or split the peak of the GW waveform \cite{Yamada:2020zvt}. By checking for waveform peak splitting in the first ever detected GW event, GW150914, Ref. \cite{Kostelecky:2016kfm} placed the first constraint on velocity birefringence, which corresponds to $|({\cal W}_1 -{\cal G}_1)/{\cal G}_0| \lesssim 10^{-11}\; {\rm km}$. Recently, the constraints from considering the width of the peak at the maximal amplitude of GW events have been improved significantly with an analysis of 50 GW events in GWTC-1 \cite{Shao:2020shv} and GWTC-2 \cite{Wang:2021ctl}. 

\begin{figure}
{
\includegraphics[width=8.1cm]{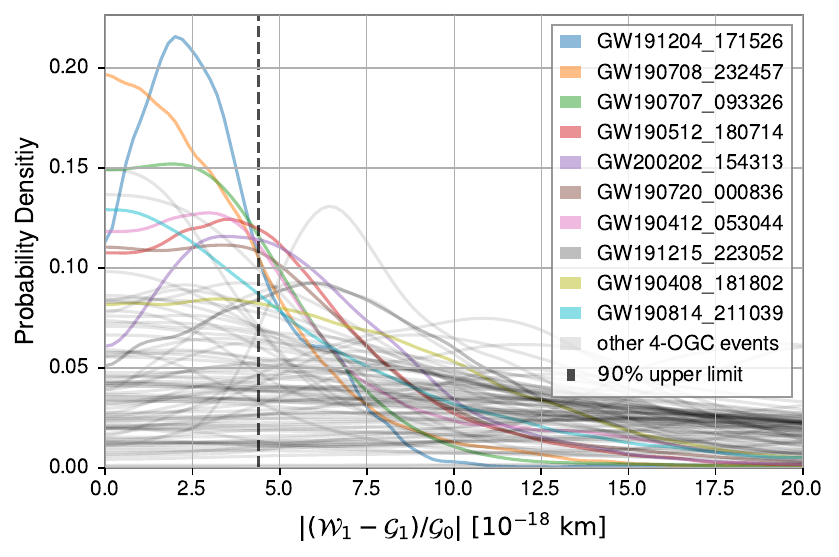}}\\
\caption{The posterior distributions for $|{\cal W}_1- {\cal G}_1|$ from 92 GW events in the 4-OGC. The legend indicates the events that give the tightest constraints. The vertical dash line denotes the 90\% upper limits from the combined result. } \label{WG2}
\end{figure}

Similar to the case of amplitude birefringence, the velocity birefringence due to parity violation can also be tested by comparing the modified waveform (\ref{gw_velocity}) with the GW strain data from the GW detectors; see Ref. \cite{Qiao:2022mln} for a review. Based on this, the tests of velocity birefringence have been carried out through full Bayesian parameter estimations on the GW events observed by the LIGO/Virgo/KAGRA detectors in a series of papers \cite{Wang:2017igw, Wang:2020cub, Wang:2021gqm, Zhao:2022pun, Wang:2020pgu, Niu:2022yhr, Gong:2021jgg, Wu:2021ndf}.  Here we would like to derive the constraint on the coefficient $({\cal W}_1 - {\cal G}_1)/{\cal G}_0$ by directly using the posterior samples obtained in \cite{Wang:2021gqm} for testing the velocity birefringent effect with 94 GW events reported in the 4th-Open Gravitational Wave Catalog (4-OGC) \cite{Nitz:2021zwj}. 
The data for these posterior samples were downloaded from \cite{web2}. With these data, one can derive the posterior distributions of $({\cal W}_1 - {\cal G}_1)/{\cal G}_0$ from the posterior distributions of the sampled parameter $M_{\rm PV}^{-1}$ for each GW event. In Fig.~\ref{WG2}, we show the posterior distributions of $|({\cal W}_1 - {\cal G}_1)/{\cal G}_0|$ for the 92 analyzed GW events \footnote{Here we exclude the two events GW190521 and GW191109 since their posterior samples show intriguing non-zero results for velocity birefringence. Some possible reasons that produce such signatures were also explored in Ref. \cite{Wang:2021gqm}. }. Note that here we treat the coefficient ${\cal W}_1 - {\cal G}_1$ as a constant as well. From this figure, we find that the posterior distributions of $|({\cal W}_1 - {\cal G}_1)/{\cal G}_0|$ with 90\% confidence intervals are consistent with the GR value ${\cal W}_1 - {\cal G}_1=0$ for all 92 GW events. By considering $|({\cal W}_1 - {\cal G}_1)/{\cal G}_0|$ as a universal parameter, we also present its upper bound from the combined posterior probability distributions in Fig.~\ref{WG2} (the vertical dashed line), from which one is able to place a constraint on $|({\cal W}_1 - {\cal G}_1)/{\cal G}_0|$ as
\bqn
|({\cal W}_1 - {\cal G}_1)/{\cal G}_0| <  4.4 \times 10^{-18} \; {\rm km}, \lb{WG44}
\eqn
at the 90\% confidence level. This bound corresponds to 
\bqn
M_{\rm Pl}^{-2}\left| \frac{1}{4} \Big(\dot c_1^{(1,3)} + \dot c_2^{(1,3)}\Big) + c_1^{(0,3)} - 3  c_{3}^{(1,3)} H  \right. \nb\\
 \left. - \frac{1}{2} \Big[c_1^{(2,1)}- (c_1^{(3,1)} - 2 c_2^{(3,1)} - 3 c_3^{(3,1)}) H \Big] \right|  \nb\\
 ~~~~~~~~~~~~~~~~~~~~~~ < 1.1 \times 10^{-18} \; {\rm km} .
\eqn
It is worth mentioning here that a slightly stronger bound on the velocity birefringence parameter was derived in Ref. \cite{Zhao:2022pun} by performing a full Bayesian analysis on GW events in the LIGO-Virgo catalog GWTC-3.

 \subsection{Lorentz-violating effects from ${\cal W}_2$ and ${\cal G}_2$}
 
The Lorentz-violating high-derivative terms introduced in Eq. (\ref{4order}) into the action of the spatial covariant gravities lead to nonzero coefficients ${\cal W}_2$ and ${\cal G}_2$. The coefficient ${\cal W}_2$ arises from a term with four spatial derivatives, and thus it modifies the usual dispersion relation of GWs in GR. The coefficient ${\cal G}_2$, which arises from a term contained two time derivatives and two spatial derivatives, not only modifies the dispersion relation but also leads to a modified damping rate of GWs during propagation \footnote{Note that in some Lorentz-violating theories, vector modes can appear and propagate in the spacetime. This kind of vector mode is generated, for example in Einstein-\AE{}ther theory, by a  time-like \ae{}ther field introduced in the theory. Such a time-like \ae{}ther field provides a preferred time direction thus breaking the Lorentz symmetry. This property is different from the spatial covariant gravities we considered here, in which only an extra scalar mode could be generated due to the breaking of the time diffeomorphism and there is no new fundamental field is introduced in the theory. }. In the following, we discuss the effects of Lorentz-violating high derivatives in the action of spatial covariant gravity on the damping rate and dispersion of GWs respectively. 
 
 \subsubsection{Lorentz-violating damping rate}
 
 The coefficient ${\cal G}_2$ induces a frequency-dependent friction term in the propagation equation of GWs. In the parametrization of Eq. (\ref{parameter_eom}), this friction term leads to a nonzero parameter $\bar \nu$, which provides a frequency-dependent damping of GW amplitudes during propagation. This implies that at different frequencies, GWs can experience different damping rates. This effect provides an amplitude modulation to the gravitational waveform \cite{Zhao:2019xmm},
 \bqn
 h_A = h_{A}^{\rm GR} \exp{\left(- \frac{1}{2}\int_{t_e}^{t_0} {\cal H} \bar \nu\right)} = h_A^{\rm GR} e^{\delta h_2}
 \eqn
 with
  \bqn
 \delta h_2 &=& - \frac{1}{2} \left.\left[a_{\bar \nu} \left(\frac{k}{aM_{\rm LV}}\right)^{\beta_{\bar \nu}}\right]\right|^{a_0}_{a_e} = - \left.\frac{1}{2} \frac{{\cal G}_2 k^2}{{\cal G}_0 a^2} \right|^{a_0}_{a_e},\nb\\
 \eqn
 where $h^{\rm GR}$ denotes the waveform of GWs in GR. We can convert the left- and right-hand GW polarization modes into the plus and cross modes, i.e., 
 \bqn
 h_{+}(f) = h_{+}^{\rm GR} e^{\delta h_2}, \\
 h_{\times}(f) = h_{\times}^{\rm GR} e^{\delta h_2}.
  \eqn
With this modified waveform, it is possible to derive the constraint on ${\cal G}_2/{\cal G}_0$ by comparing the modified waveform with the GW strain data from the GW detectors. However, no test on the frequency-dependent damping effects has not been carried out yet in the literature, and we expect to consider this in our future works.
 

 \subsubsection{Lorentz-violating dispersion relation}
 
 Lorentz violation of gravity in general modifies the conventional linear dispersion relation to a nonlinear one. Due to the existence of the coefficients ${\cal W}_2$ and ${\cal G}_2$ in spatial covariant gravity, the dispersion of GWs becomes
 \bqn
 \omega_k^2 = k^2(1+ \bar \mu)
 \eqn
 with 
 \bqn
 \bar \mu = \frac{{\cal W}_2 -{\cal G}_2}{{\cal G}_0} \frac{k^2}{a^2}.
 \eqn
With this modified dispersion relation, the phase velocity of GWs reads
\bqn
v &\simeq& 1 - \frac{1}{2} \alpha_{\bar \mu} \left(\frac{k}{a M_{\rm LV}}\right)^{\beta_{\bar \mu}}  \nb\\
&=&1- \frac{1}{2}\frac{{\cal W}_2 -{\cal G}_2}{{\cal G}_0} \frac{k^2}{a^2}.
\eqn
Consider GWs emitted at two different times $t_e$ and $t_e'$, with wave number $k$ and $k'$, and  received at corresponding arrival times $t_0$ and $t_0'$;  then, the different velocities of modes lead to a difference in their arrival times \cite{Zhao:2019xmm, Mirshekari:2011yq},
\bqn
\Delta t_0 &=& (1+z) \Delta t_e + \frac{1}{2} \left(\frac{k^{\beta_{\bar \mu}}}{M_{\rm LV}^{\beta_{\bar \mu}}} -\frac{k'^{\beta_{\bar \mu}}}{M_{\rm LV}^{\beta_{\bar \mu}}} \right)\int_{t_e}^{t_0} \frac{\alpha_{\bar \mu}}{a^{\beta_{\bar \mu}+1}} dt\nb\\
&=&(1+z)\Delta t_e + \frac{k^2-k'^2}{2} \int_{t_e}^{t_0} \frac{{\cal W}_2 - {\cal G}_2}{{\cal G}_0a^{3}}dt.
\eqn
Here $\Delta t_e = t_e -t_e'$. This velocity difference induces a modification of the phase of the GW signal emitted from a binary compact star system. The modified GW waveform in the Fourier domain reads  \cite{Zhao:2019xmm, Mirshekari:2011yq}
\bqn
h_A(f) = h_A^{\rm GR} e^{i \delta \Psi_2}, \lb{gw_lorentz}
\eqn
where the phase correction $\delta \Psi_2$ induced by the velocity difference is expressed as
\bqn
\delta \Psi_2 = A_\mu (\pi f)^3
\eqn
with
\bqn
A_{\bar \mu} &=&\frac{4}{3} \int_{t_e}^{t_0} \frac{{\cal W}_2 -{\cal G}_2}{{\cal G}_0a^3} dt \nb\\
&=& \frac{4}{3H_0} \int_0^z \frac{({\cal W}_2 - {\cal G}_2)(1+z')^2dz'}{{\cal G}_0\sqrt{\Omega_m(1+z')^3+\Omega_\Lambda}}. \lb{Amu}
\eqn

We analyze the GW constraints on the Lorentz-violating dispersion relation by comparing the modified waveform (\ref{gw_lorentz}) with the GW strain data in GWTC-1 \cite{testGR_GWTC1}, GWTC-2 \cite{testGR_GWTC2}, and GWTC-3 \cite{LIGOScientific:2021sio}.  The gravitational constraint on $({\cal W}_2 - {\cal G}_2)/{\cal G}_0$ can be obtained from the posterior samples of the Lorentz-violating parameter $A_{4}$ in Refs. \cite{testGR_GWTC1, testGR_GWTC2, LIGOScientific:2021sio}. In Refs. \cite{testGR_GWTC1, testGR_GWTC2, LIGOScientific:2021sio}, the Lorentz-violating parameter $A_4$ was sampled separately for $A_4>0$ and $A_4<0$ in Refs. \cite{testGR_GWTC1, testGR_GWTC2, LIGOScientific:2021sio}. Here we consider positive and negative $A_4$  separately as well and derive the corresponding bounds on $({\cal W}_2 - {\cal G}_2)/{\cal G}_0$, which are presented in Table \ref{tableGW}. From this table, we see that the most stringent constraint is from the combined posterior of GW events in GWTC-3, which gives
\bqn
| ({\cal W}_2- {\cal G}_2)/{\cal G}_0 |< 1.2 \times 10^{-10} \; {\rm m}^2 \nb\\
\eqn
at 90\% C.L. This bound corresponds to 
\bqn
M_{\rm Pl}^{-2}\left| c_{3}^{(0, 4)} + c_1^{(2, 2)} \right| < 6 \times 10^{-11} \; {\rm m}^2.
\eqn
 
\begin{table}
\caption{90\% confidence level upper bounds on $|({\cal W}_2 - {\cal G}_2)/{\cal G}_0|$ for positive and negative ${\cal W}_2 - {\cal G}_2$ respectively from the Bayesian inference by analyzing GW events in the LIGO/Virgo/KAGRA catalogs GWTC-1 \cite{testGR_GWTC1}, GWTC-2 \cite{testGR_GWTC2}, and GWTC-3 \cite{LIGOScientific:2021sio}. Note that the bounds on $|({\cal W}_2 - {\cal G}_2)/{\cal G}_0|$ is in the unit of $10^{-10} \; m^2$ with $m$ being the meter.}
\lb{tableGW}
\begin{ruledtabular}
\begin{tabular} {c|cc}
catalogs & \makecell{sampled with \\ negative ${\cal W}_2 - {\cal G}_ 2$} &  \makecell{sampled with \\positive ${\cal W}_2 - {\cal G}_ 2$} \\
\hline 
GWTC-1
& 9.0 &  5.5 \\
GWTC-2
& 3.4 &  2.5 \\
GWTC-3
& 2.9 &  1.2 \\
\end{tabular}
\end{ruledtabular}
\end{table}
 
 \begin{table*}
\caption{Summary of estimations for bounds of the coupling coefficients in spatial covariant gravities. Note that all the coefficients are estimated approximately at present time, i..e, $z=0$. Here $[a_{\rm min}, b_{\rm max}]$ represents constraints with $a_{\rm bin}$ and $b_{\rm max}$ being the lower the upper bounds, respectively.}
\lb{table4}
\begin{ruledtabular}
\begin{tabular} {cccc}
 Coefficients & bounds & related coefficients   & datasets used \\
 \hline
$\sqrt{\frac{{\cal W}_0}{{\cal G}_0}} -1$ & $[-30, 7] \times 10^{-16}$ & \makecell{$c_1^{(2,0)}, c_1^{(3,0)}, c_2^{(3,0)}, c_1^{(4,0)}, c_2^{(4,0)}, c_3^{(4,0)},$ \\ $c_1^{(0,2)}, c_3^{(1,2)}, c_4^{(1,2)}, c_6^{(2,2)}, c_7^{(2,2)},c_8^{(2,2)}, c_9^{(2,2)}$ }& \makecell{mult-messenge observations \\ of GW170817 \cite{LIGOScientific:2017vwq, LIGOScientific:2017zic} }\\
\hline
$\frac{(\ln{\cal G}_0)'}{{\cal H}} $ & $ [-3.0,\; 2.5]$ & $c_1^{(2,0)}, c_1^{(3,0)}, c_2^{(3,0)}, c_1^{(4,0)}, c_2^{(4,0)}, c_3^{(4,0)}$ & \makecell{dark sirens in GWTC-3 with\\  BBH mass distributions \cite{Mancarella:2021ecn} } \\
\hline
$|{\cal G}_1/{\cal G}_0|$ &\makecell{ $ \lesssim 2065 \;  {\rm km}$ \\~\\ $ \lesssim 1000 \; {\rm km} $} & $c_1^{(2,1)}, c_1^{(3,1)}, c_2^{(3,1)}, c_3^{(3,1)} $ & \makecell{tests of amplitude birefringence \\ with LIGO-Virgo O1/O2 \cite{Wang:2020cub} \\ \hline from statistic distribution of \\ $\cos \iota$ in GWTC-2 \cite{Okounkova:2021xjv}}  \\
\hline
$\left|\frac{{\cal W}_1 - {\cal G}_1}{{\cal G}_0}\right|$ & $\lesssim 4.4 \times 10^{-18} \; {\rm km}$ & \makecell{ $ c_1^{(0,3)}, c_1^{(1,3)}, c_2^{(1,3)}, c_3^{(1,3)} ,$ \\ $c_1^{(2,1)}, c_1^{(3,1)}, c_2^{(3,1)}, c_3^{(3,1)}$ } &  \makecell{tests of velocity birefringence \\with 4-OGC \cite{Wang:2021gqm}} \\
\hline
$\left|\frac{{\cal W}_2 - {\cal G}_2}{{\cal G}_0}\right|$ & $\lesssim 1.2 \times 10^{-10} \; {\rm m}^2$  & $c_1^{(2, 2)}, \; c_3^{(0, 4)}$ & \makecell{tests of Lorentz-violating \\dispersion with GWTC-3 \cite{LIGOScientific:2021sio}} \\
\end{tabular}
\end{ruledtabular}
\end{table*}

\section{Summary and Discussions}
 \renewcommand{\theequation}{6.\arabic{equation}} \setcounter{equation}{0}

The spatial covariant gravity is only invariant under the three-dimensional spatial diffeomorphism, which breaks the time diffeomorphism.  Therefore the gravitational action of this type of theory can be constructed in terms of spatial diffeomorphism invariants. A lot of scalar-tensor theories can be mapped to the spatial covariant framework by imposing the unitary gauge on the coupling scalar field. This provides us with a general framework for exploring the effects of unknown high-energy physics on the propagation of GWs. 

In this paper, we studied the effects of the spatial covariant gravities on the propagation of GWs, produced by the coalescence of compact binaries, and their observational constraints with GW events from the LIGO/Virgo/KAGRA Collaboration. For this purpose, we calculated the effects of the spatial covariant gravities on the friction, speed, amplitude, and velocity birefringences of GWs, as well as the modified dispersion relation during GW propagation in the cosmological background. These effects can be described by the universal parametrization proposed in Refs. \cite{Zhao:2019xmm, Qiao:2019wsh}. Different effects correspond to different parameters, as discussed in detail in Sec. III and summarized in Table \ref{paramaters}. These effects can be divided into three classes: 1) frequency-independent effects which include modifications to GW speed and friction; (2) parity-violating effects which include the amplitude and velocity birefringences of GWs; and 3) Lorentz-violating effects which include the modified damping rate and dispersion relation of GWs. Among these effects, the parity-violating and Lorentz-violating effects are frequency dependent. Depending on different coefficients in the spatial covariant gravities, these frequency-dependent effects can produce amplitude modulation and phase corrections in the waveform of GWs produced by the coalescence of compact binaries. The calculation of these modified GW waveforms was presented in Sec.~III. These modified waveforms provide important tools for constraining parity-violating or Lorentz-violating effects in the spatial covariant gravities with current available GW events or future GW detections. 

We compared these new effects with the publicly available posterior samples or results from various tests of gravities using LIGO/Virgo/KAGRA data to obtain constraints on coupling coefficients in the action of spatial covariant gravities. Different effects can be tested with different phenomena in GW observations. For the frequency-independent effects, GW speed is constrained by the multimessenger observation of GW170817/GRB170817A, while GW friction is constrained by dark sirens in GWTC-1 with BBH population models. For the parity-violating effects, we constrained the amplitude and velocity birefringent parameters from full Bayesian parameter estimations of the GW events observed by the LIGO/Virgo/KAGRA detectors. We also reported the constraint on the amplitude birefringent parameter from the statistic analysis of the posterior distributions of $\cos \iota$ for GW events in GWTC-2. For the Lorentz-violating effects, we reported the constraint on the parameter in the modified dispersion relation from a Bayesian analysis of GW events in GWTC-3. Our results are summarized in Table \ref{table4}. Using these constraints, we also derived the corresponding bounds on the coupling coefficients of a number of specific theories in the Appendix.

It is remarkable that the constraints on the effects that modify the speed of GWs (including the effects of frequency-independent modification to GW speed, amplitude birefringence, and Lorentz-violating modified dispersion) are more stringent than those that affect the amplitude of GWs. For example, for parity-violating effects, the parameter $|{\cal G}_1|$ (with ${\cal W}_1 =0$) can be constrained from tests of amplitude or velocity birefringence. It is evident from Table \ref{table4} that the constraint on $|{\cal G}_1|$ from the tests of velocity birefringence is stronger than that from tests of amplitude birefringence by 20 orders of magnitude. This is because ground-based detectors are more sensitive to phase corrections than amplitude modulations for the tests involving GW signals from the coalescence of compact binaries.

Here we would like to mention that the study performed in this paper can be extended in a few directions in future works. First, in order to derive the bounds on the coupling coefficients, we considered the different effects of spatial covariant gravities on the propagation of GWs separately. Thus, it would be interesting to consider all of the new effects that arise from the different coefficients together. To do this, one would need to simulate the modified waveform with GW data by sampling all of the relevant coefficients. Second, it would be interesting to see if future GW detectors such as the third-generation ground-based detectors, space-based detectors, and pulsar-timing arrays, can improve the bounds obtained in this paper. For the spatial covariant gravities, most of the effects on the propagation of GWs are very sensitive to the higher frequency of GWs. This is because the amplitude and phase corrections to the waveform are proportional to $f^{\beta_{\bar \nu, \nu_A}}$ and  $f^{1+\beta_{\bar \mu, \mu_A}}$, respectively. For this reason, it is not likely that future space-based detectors or pulsar-timing arrays will be able to improve the bounds given in this paper since the sensitive frequency of these detectors is much lower than those of the ground-based detectors.  For the velocity birefringence effect,  it was shown \cite{Wang:2020cub} that the third-generation gravitational-wave detectors are able to improve the constraint on the energy scale of parity violation to ${\cal O}(10^2) \; {\rm GeV}$. This implies that the bound on $|({\cal W}_2-{\cal G}_2)/{\cal G}_0|$ can be improved by about 3 orders of magnitude, i.e., improved to be $|({\cal W}_2-{\cal G}_2)/{\cal G}_0| \lesssim {\cal O}(10^{-21})\; {\rm km}$. Similarly, it is expected that the third-generation gravitational-wave detectors could be able to significantly improve the constraints on the other effects, such as the amplitude birefringence, Lorentz-violating damping rate, and modified dispersion relations. We expect to come back to these issues soon in future work.

\section*{Acknowledgements}
T.Z. and A.W. are supported in part by the National Key Research and Development Program of China Grant No.2020YFC2201503, and the Zhejiang Provincial Natural Science Foundation of China under Grant No. LR21A050001 and LY20A050002, the National Natural Science Foundation of China under Grant No. 12275238, No. 11975203, No. 11675143, and the Fundamental Research Funds for the Provincial Universities of Zhejiang in China under Grant No. RF-A2019015. W.Z. is supported by the National Key Research and Development Program of China Grant No.2021YFC2203102 and 2022YFC2200100, NSFC Grants No. 12273035 and 11903030, the Fundamental Research Funds for the Central Universities. 

\appendix

\section*{Appendix A: Several specific spatial covariant gravities}
 \renewcommand{\theequation}{A.\arabic{equation}} \setcounter{equation}{0}

The spatial covariant gravities provide a unifying framework for describing scalar-tensor theories in the unitary gauge. In this appendix, we present several specific scalar-tensor theories in the unitary gauge and Lorentz-violating gravity by writing their gravitational actions in the form of Eq. (\ref{oldmodel}). We also provide relations between the coupling coefficients in each theory and the corresponding coefficients in the spatial covariant gravities. In addition, the observational constraints on these theories are derived from the constraints presented in Table \ref{table4}.

\subsection{Horndeski theory}

The Horndeski theory is a general scalar-tensor theory constructed from the metric tensor $g_{\mu\nu}$ and a scalar field $\phi$ and can have a second-order field equation \cite{Horndeski:1974wa}. The Lagrangian of the Horndeski theory in the unitary gauge with $\phi=\phi(t)$ can be found in Refs. \cite{Gleyzes:2013ooa, Fujita:2015ymn, Gao:2019liu}, and is written as \cite{Gao:2019liu}
\bqn
{\cal L}_{\rm H}^{\rm unitary} &=& a_0 K - 2 a_1 (R_{ij}-\frac{1}{2}R g_{ij}) K^{ij} \nb\\
&&+ b_0 (K_{ij}K^{ij} - K^2)  \nb\\
&&+ c_0 (K^3 - 3 K K_{ij}K^{ij} + 2 K^i_j K^i_k K^k_i) \nb\\
&&+ d_0 + d_1 R,
\eqn
where the six coefficients $a_0, a_1, b_0, c_0, d_0, d_1$ are functions of $t$ and $N$, which can be related to the coefficients of the  Horndeski theory through Eqs.~(8)-(13) in Ref. \cite{Gao:2019liu}. Comparing the above Lagrangian with Eq. (\ref{oldmodel}), one finds
\bqn
&&c_{1}^{(0, 0)} = d_0, \;\; c_1^{(1, 0)} = a_0, \;\; c_1^{(0, 2)}=d_1, \;\; c_1^{(2, 0)} = b_0, \nb\\
&&c_2^{(2,0)} = - b_0, \;\;  c_1^{(3,0)} = 2c_0,\;\; c_2^{(3, 0)} = -3 c_0, \;\; c_3^{(3,0)}=c_0, \nb\\
&&c_3^{(1,2)} = -2 a_1, \;\; c_4^{(1,2)} = a_1,
\eqn
with all other coefficients $c_i^{(t, s)}=0$. The corresponding coefficients ${\cal G}_0, {\cal G}_1, {\cal G}_2$ and ${\cal W}_0, {\cal W}_1, {\cal W}_2$ in the GW propagation equation (\ref{eom_A}) are 
\bqn
{\cal G}_0 = \frac{1}{2}b_0- \frac{3}{2}c_0 H, \;\; {\cal G}_1=0={\cal G}_2, \\
{\cal W}_0=\frac{1}{2}d_1 - \frac{1}{2}\dot a_1, \;\; {\cal W}_1=0={\cal G}_2.
\eqn
Then using the constraints in Table.~\ref{table4}, it is easy to infer that
\bqn
-3 \times 10^{-15} <\sqrt{\frac{d_1-\dot a_1}{b_0-3c_0 H}}-1 < 7 \times 10^{-16}
\eqn
from the multimessenger observations of GW170817/GRB170817A \cite{LIGOScientific:2017vwq, LIGOScientific:2017zic}, and 
\bqn
-3 < \frac{{\dot b}_0 - 3 {\dot c}_0 H - 3 c_0 \dot H}{b_0 H - 3 c_0 H^2} < 2.5
\eqn
by using the dark sirens in GWTC-3 with BBH mass distributions \cite{Mancarella:2021ecn}.

\subsection{Scalar-Gauss-Bonnet gravity}

It is also interesting to note that  scalar-Gauss-Bonnet gravity with the Lagrangian ${\cal L}_{\rm GB} = \xi(\phi) R_{\rm GB}^2$ can be recast in the form of the Horndeski theory \cite{Kobayashi:2011nu}, where $R_{\rm GB}^2 \equiv \frac{1}{4} \varepsilon_{\mu \nu \alpha \beta} \varepsilon_{\rho \sigma}^{\lambda \tau} \;^4R^{\alpha \beta}{ }_{\lambda \tau}{ }^4R^{\mu \nu \rho \sigma}$. Here $\varepsilon^{\mu\nu\rho\sigma} $ is the Levi-Civita tensor and $\;^4R_{\rho\sigma \alpha \beta}$ is the Riemann tensor defined in four-dimensional spacetime. In the unitary gauge, the propagation equation of GWs in scalar-Gauss-Bonnet gravity corresponds to \cite{Gao:2019liu}
\bqn
{\cal G}_0 =\frac{1}{4}M_{\rm Pl}^2+2 \dot{\xi} H, \;\; {\cal G}_1=0={\cal G}_2, \\
{\cal W}_0=\frac{1}{4}M_{\rm Pl}^2+2 \ddot{\xi}, \;\; {\cal W}_1=0={\cal G}_2.
\eqn
Using the constraints in Table \ref{table4}, one obtains
\bqn
-3 \times 10^{-15}\lesssim 8\frac{\ddot{\xi}-\dot{\xi} H}{M_{\rm Pl}^2} \lesssim 7 \times 10^{-16}
\eqn
from the multimessenger observations of GW170817/GRB170817A \cite{LIGOScientific:2017vwq, LIGOScientific:2017zic}, and 
\bqn
-3 \lesssim 8 \frac{\ddot{\xi} H + \dot{\xi} \dot{H}}{M_{\rm Pl}^2 H} \lesssim 2.5
\eqn
using the dark sirens in GWTC-3 with BBH mass distributions \cite{Mancarella:2021ecn}.

\subsection{Chern-Simons gravity}

Chern-Simons gravity is an effective extension of GR that includes a coupling between a scalar field $\phi$ and the Chern-Pontryagin term \cite{Jackiw:2003pm, Lue:1998mq, Alexander:2009tp}. The Lagrangian of the new term beyond GR is
\bqn
{\cal L}_{\rm CS} = \frac{M_{\rm Pl}^2}{2}\frac{1}{8}\vartheta(\phi)\varepsilon^{\mu\nu\rho\sigma} \;^4R_{\rho\sigma \alpha \beta} \;^4R^{\alpha \beta}_{\;\; \mu\nu},
\eqn
In the unitary gauge, one can find that \cite{Gao:2019liu}
\bqn
 \mathcal{L}_{\mathrm{CS}}^{(\text {u.g.) }}&= & \frac{M_{\rm Pl}^2}{2} \varepsilon^{i j k} \vartheta \left(K_{i l} K^{l m} \nabla_j K_{k m}+K_i^l K_j^m \nabla_m K_{k l}\right. \nb\\
&& -K K_i^l \nabla_j K_{k l}-2 R_i^l \nabla_j K_{k l}-\frac{1}{N} \frac{\dot{\vartheta}}{\vartheta} K_i^l \nabla_j K_{k l} \nb\\
&& \left.-\frac{2}{N} K_i^l K_{j l} K_{k m} \nabla^m N-\frac{2}{N} \nabla_i K_{j l} \nabla_k \nabla^l N\right).\nb\\
\eqn
Ignoring the terms containing spatial derivatives of $N$ and comparing  $\mathcal{L}_{\mathrm{CS}}^{(\text {u.g.) }}$ with Eq. (\ref{oldmodel}), one has
\bqn
&&c_{1}^{(2,0)} = c_1^{(0,2)}=- c_2^{(2, 0)}=\frac{M_{\rm Pl}^2}{2} ,\nb\\
&&c_2^{(3, 1)} = - c_1^{(3, 1)}=-c_3^{(3,1)}=- \frac{1}{2} c_{1}^{(1, 3)}= \frac{M_{\rm Pl}^2}{2}\vartheta , \nb\\
&&c_1^{(2, 1)} = -\frac{M_{\rm Pl}^2}{2}\frac{\dot \vartheta}{N}.
\eqn
Then the coefficients in the propagation equation (\ref{eom_A}) read
\bqn
{\cal G}_0 =\frac{1}{4}M_{\rm Pl}^2, \;\; {\cal G}_1=- \frac{M_{\rm Pl}^2}{2} \frac{\dot \vartheta}{2}, \;\; {\cal G}_2=0, \\
{\cal W}_0=\frac{1}{4}M_{\rm Pl}^2, \;\; {\cal W}_1=-\frac{M_{\rm Pl}^2}{2}  \frac{\dot \vartheta}{2}, \;\; {\cal G}_2=0.
\eqn
The nonzero coefficient ${\cal G}_1$ induces amplitude birefringence in the propagation of GWs. Using the constraints in Table.~\ref{table4}, one gets
\bqn
|\dot \vartheta| \lesssim 2065 \; {\rm km}
\eqn
from tests of amplitude birefringence with LIGO-Virgo O1/O2 events \cite{Wang:2020cub} and
\bqn
|\dot \vartheta| \lesssim 1000 \; {\rm km}
\eqn
from the analysis of statistic distribution of $\cos \iota$ in GWTC-2 \cite{Okounkova:2021xjv}.

\subsection{Lorentz-violating Weyl gravity}

Weyl gravity modifies GR by adding a Weyl-squared term to the gravitational Lagrangian \cite{Deruelle:2012xv}, 
\bqn
{\cal L}_{\rm Weyl} = -\frac{M_{\rm Pl}^2}{2} \frac{\gamma}{2} \;^4C_{\mu\nu\rho\lambda} \;^4C^{\mu\nu\rho\lambda},
\eqn
where $\;^4C_{\mu\nu\rho\lambda}$ is the Weyl tensor in four-dimensional spacetime. In the unitary gauge, the Lagrangian ${\cal L}_{\rm Weyl}$ becomes \cite{Gao:2019liu, Deruelle:2012xv}
\bqn
\mathcal{L}_{\mathrm{DSSY}}^{\text {(u.g.) }} &= & M_{\rm Pl}^2 \gamma( -\nabla_i K \nabla^i K-\nabla_i K^{i k} \nabla_j K_k^j+2 \nabla^i K \nabla_j K_i^j \nb\\
&& -2 \nabla_k K_{i j} \nabla^j K^{i k}+2 \nabla_k K_{i j} \nabla^k K^{i j}).
\eqn
Comparing this with Eq. (\ref{oldmodel}), we have
\bqn
&&c_{1}^{(2,0)} = c_1^{(0,2)}=- c_2^{(2, 0)}=\frac{M_{\rm Pl}^2}{2} ,\nb\\
&&c_1^{(2, 2)}=-c_2^{(2, 2)}=c_4^{(2, 2)}= -\frac{1}{2} c_3^{(2, 2)}=-\frac{1}{2} c_5^{(2, 2)} = 2M_{\rm Pl}^2 \gamma .\nb\\
\eqn
Then, the coefficients in the propagation equation (\ref{eom_A}) read
\bqn
{\cal G}_0 =\frac{1}{4}M_{\rm Pl}^2, \;\; {\cal G}_1=0, \;\; {\cal G}_2= M_{\rm Pl}^2 \gamma, \\
{\cal W}_0=\frac{1}{4}M_{\rm Pl}^2, \;\; {\cal W}_1=0, \;\; {\cal G}_2=0.
\eqn
Using the constraints in Table.~\ref{table4}, one gets
\bqn
|4\gamma| \lesssim 1.2 \times 10^{-10} \; {\rm m}^2
\eqn
from tests of Lorentz-violating dispersion with GWTC-3 \cite{LIGOScientific:2021sio}.

\subsection{Chiral scalar-tensor theory}

Chern-Simons gravity can be extended to include parity-violating curvature terms with couplings between the Riemann curvature and derivatives of the scalar field. One theory of this type is the chiral scalar-tensor theory considered in Ref. \cite{Crisostomi:2017ugk}, which includes couplings between the Riemann curvature and the first and second derivatives of the scalar field. 

The coupling of the first derivatives of the scalar field contains four terms with coupling coefficients $(a_1, a_2, a_3, a_4)$ in the Lagrangian \cite{Crisostomi:2017ugk}, which can be written in the unitary gauge as
\bqn
{\cal L}_{\rm chiral}^{\rm u.g.1} &=&M_{\rm Pl}^2 \frac{\dot \phi^2}{N^2}\Big[(4a_1 + a_3) \varepsilon_{i j k} K^{l i} K^{m j} \nabla_m K_l^k \nb\\
&&- (4a_1 + 2a_3) \varepsilon_{i j k} R^{l i} \nabla^k K_l^j\nb\\
&& + a_3 \varepsilon_{i j k}\left(K_m^i K^{l m}-K K^{l i}\right) \nabla^k K_l^j\Big].
\eqn
In the above, we have used the condition $4 a_1+2 a_2+a_3+8 a_4=0$ which makes the theory to be healthy in the unitary gauge. 

For the coupling between the Riemann curvature and the second derivatives of the scalar field, there are seven terms with coupling coefficients $(b_1, b_2, b_3, b_4, b_5, b_6, b_7)$ in the Lagrangian \cite{Crisostomi:2017ugk}, which can be written in the unitary gauge as
\bqn
{\cal L}_{\rm chiral}^{\rm u.g.2}&=&M_{\rm Pl}^2 \frac{{\dot \phi}^3}{N^3} b_1 \varepsilon_{i j k} K^{l i} \nabla^k K_l^j \nb\\
&&+M_{\rm Pl}^2 \frac{\dot{\phi}^4}{N^4} (b_4+b_5-b_3) \varepsilon_{i j k} K^{l i} K^{m j} \nabla_m K_l^k.\nb\\
\eqn
In the above, we dropped all of the terms containing spatial derivatives of the lapse function, since they do not contribute to the propagation of the tensorial GWs, and we used the conditions $b_7=0$, $b_6=2(b_4+b_5)$, and $b_2=- \frac{\dot \phi^2}{2N^2} (b_3-b_4)$ to make the theory to be healthy when the unitary gauge is imposed \cite{Crisostomi:2017ugk}. 

Now the parity-violating Lagrangian of the chiral scalar-tensor theory reads ${\cal L}^{\rm u.g.}_{\rm chiral}= {\cal L}_{\rm CS}^{\rm u.g.} + {\cal L}_{\rm chiral}^{\rm u.g.1}+ {\cal L}_{\rm chiral}^{\rm u.g.2}$. Considering it with Eq. (\ref{oldmodel}), one obtains
\bqn
&&c_{1}^{(2,0)} = c_1^{(0,2)}=- c_2^{(2, 0)}=\frac{M_{\rm Pl}^2}{2} ,\nb\\
&& c_1^{(3, 1)} = \frac{M_{\rm Pl}^2}{2}\Big[- \vartheta - \frac{2 {\dot \phi}^2}{N^2}(4a_1+a_3)  \nb\\
&&~~~~~~~~ - \frac{2 {\dot \phi}^4}{N^4}(b_4+b_5-b_3)\Big], \nb\\
&&c_{1}^{(1, 3)}= \frac{M_{\rm Pl}^2}{2}\Big[- 2 \vartheta+\frac{2 {\dot \phi}^2}{N^2}(4a_1+2a_3)\Big],\nb\\
&& c_1^{(2, 1)} =\frac{M_{\rm Pl}^2}{2}\Big[- -\frac{\dot \vartheta}{N}- \frac{2 \dot \phi^3}{N^3}b_1\Big], \nb\\
&&c_2^{(3, 1)}  = \frac{M_{\rm Pl}^2}{2}\Big[-\vartheta - \frac{2 \dot \phi^2}{N^2}a_3\Big], \nb\\
&& c_3^{(3, 1)} = \frac{M_{\rm Pl}^2}{2} \Big[-  \vartheta +\frac{2 \dot \phi^2}{N^2}a_3\Big].
\eqn 
Then the coefficients in the propagation equation (\ref{eom_A}) read
\bqn
{\cal G}_0 &=&\frac{1}{4}M_{\rm Pl}^2, \;\;  {\cal G}_2=0, \nb\\
{\cal G}_1&=&  \frac{M_{\rm Pl}^2}{2} \Big[- \frac{\dot \vartheta}{2} -b_1 \dot \phi^3 +2 (a_3+2a_1) \dot \phi^2 H \nb\\
&&+ (b_4+b_5-b_3) \dot \phi^4 H\Big],  \nb\\
{\cal W}_0&=&\frac{1}{4}M_{\rm Pl}^2, \;\;  {\cal G}_2=0, \nb\\
{\cal W}_1&=& \frac{M_{\rm Pl}^2}{2} \Big[- \frac{\dot \vartheta}{2} +  (4a_1+2a_3)\dot \phi \ddot \phi +(2 \dot a_1+\dot a_3) \dot \phi^2\Big]. \nb\\
\eqn
The nonzero coefficient ${\cal G}_1$ induces amplitude birefringence in the propagation of GWs and the nonzero ${\cal G}_1- {\cal W}_1$ leads to velocity birefringence. For $|{\cal G}_1/{\cal G}_0|$, using the constraints in Table.~\ref{table4}, one gets
\bqn
&& \Big|- \dot \vartheta -2b_1 \dot \phi^3 +4(a_3+2a_1) \dot \phi^2 H  \nb\\
&&~~~~ + 2 (b_4+b_5-b_3) \dot \phi^4 H \Big| \lesssim 2065 \; {\rm km}
\eqn
from tests of amplitude birefringence with LIGO-Virgo O1/O2 events \cite{Wang:2020cub} and
\bqn
&& \Big|- \dot \vartheta -2b_1 \dot \phi^3 +4(a_3+2a_1) \dot \phi^2 H  \nb\\
&&~~~~ + 2 (b_4+b_5-b_3) \dot \phi^4 H \Big| \lesssim 1000 \; {\rm km}
\eqn
from the analysis of statistic distribution of $\cos \iota$ in GWTC-2 \cite{Okounkova:2021xjv}. For $({\cal W}_1 - {\cal G}_1)/{\cal G}_0$, one has
\bqn
&& \Big| (a_3+2a_1) \dot \phi^2 H -(4a_1+2a_3)\dot \phi \ddot \phi -(2 \dot a_1+\dot a_3) \dot \phi^2 \nb\\
&&~ - b_1 \dot \phi^3 + (b_4+b_5-b_3) \dot \phi^4 H \Big| \lesssim 2.2 \times 10^{-18}\; {\rm km}, \nb\\
\eqn
from the tests of velocity birefringence with 4-OGC in Ref. \cite{Wang:2021gqm}.

\end{document}